%% file: hd32115-v5.0.tex
\documentclass[useAMS,usenatbib]{mn2e}
\usepackage{txfonts}
\usepackage{longtable}  
\usepackage{rotating}
\usepackage{natbib}
\usepackage{graphicx}
\usepackage{graphics}
\usepackage{psfrag}
\usepackage{amssymb}
\usepackage{multicol}
\bibpunct{(}{)}{;}{a}{}{,}
\def\Teff{\ensuremath{T_{\mathrm{eff}}}}
\def\logg{\ensuremath{\log g}}

\def\vmic{$\upsilon_{\mathrm{mic}}$}
\def\vmac{$\upsilon_{\mathrm{macro}}$}
\def\vsini{\ensuremath{{\upsilon}\sin i}}
\def\kms{$\mathrm{km\,s}^{-1}$}

\def\exc{$\chi_{\mathrm{excit}}$}
\def\loggf{$\log{gf}$}

\def\nlte{non-LTE}
\def\llm{{\sc LLmodels}}

\def\M{\ensuremath{M/M_{\odot}}}
\def\R{\ensuremath{R/R_{\odot}}}
\def\vald{{\sc VALD}}
\def\synth{{\sc SYNTH3}}
\title[The accuracy of stellar atmospheric parameter determinations: a case study with HD\,32115 and HD\,37594]
{The accuracy of stellar atmospheric parameter determinations: a case study with HD\,32115 and HD\,37594\thanks{Data
obtained with the 2.7-m telescope at McDonald Observatory, Texas, US.}} 

\author[L. Fossati et al.]{L. Fossati,$^{1}$\thanks{l.fossati@open.ac.uk}
			   T. Ryabchikova,$^{2}$
			   D.\,V. Shulyak,$^{3}$
			   C.\,A. Haswell,$^{1}$
			   A. Elmasli,$^{4}$
			   C.\,P. Pandey,$^{5}$
		\newauthor T.\,G. Barnes,$^{6}$
			   K. Zwintz$^{7}$\\
$^{1}$Department of Physics and Astronomy, Open University, 
Walton Hall, Milton Keynes MK7 6AA, UK\\
$^{2}$Institute of Astronomy, Russian Academy of Sciences, Pyatnitskaya 48, 
119017 Moscow, Russia\\
$^{3}$Institute of Astrophysics, Georg-August-University, 
Friedrich-Hund-Platz 1, D-37077, G\"ottingen, Germany\\
$^{4}$Department of Astronomy and Space Sciences, Ankara University, 06100, 
Tando\u{g}an, Ankara, Turkey\\
$^{5}$Aryabhatta Research Institute of Observational Sciences, Nainital,
263129, India\\
$^{6}$The University of Texas at Austin, McDonald Observatory, 1 University 
Station, C1402, Austin, Texas, 78712-0259, USA\\
$^{7}$Institut f\"ur Astronomie, Universit\"{a}t Wien, 
T\"{u}rkenschanzstrasse 17, 1180 Wien, Austria}
\begin{document}

\date{}

\pagerange{\pageref{firstpage}--\pageref{lastpage}} \pubyear{2011}

\maketitle

\label{firstpage}

\begin{abstract}
We present detailed parameter determinations of two chemically normal late 
A-type stars, HD\,32115 and HD\,37594, to uncover the reasons behind large 
discrepancies between two previous analyses of these stars performed with 
a semi-automatic procedure and a ``classical" analysis. Our study is based 
on high resolution, high signal-to-noise spectra obtained at the McDonald 
Observatory. Our method is based on the simultaneous use of all available 
observables: multicolor photometry, pressure-sensitive magnesium lines, 
metallic lines and Balmer line profiles. Our final set of fundamental 
parameters fits, within the error bars, all available observables. It differs
from the published results obtained with a semi-automatic procedure. A direct 
comparison between our new observational material and the spectra previously 
used by other authors shows that the quality of the data is not the origin 
of the discrepancies. As the two stars require a substantial 
macroturbulence velocity to fit the line profiles, we concluded that 
neglecting this additional broadening in the semi-automatic analysis is 
one origin of discrepancy. The use of Fe\,{\sc i} excitation equilibrium 
and of the Fe ionisation equilibrium, to derive effective temperature and 
surface gravity, respectively, neglecting all other indicators leads to a 
systematically erroneously high \Teff. We deduce that the results 
obtained using only one parameter indicator might be biased and that those
results need to be cautiously taken when performing further detailed analyses,
such as modelling of the asteroseismic frequencies or characterising
transiting exoplanets.
\end{abstract}

\begin{keywords}
techniques: spectroscopic -- stars: fundamental parameters -- 
stars: individual: HD\,32115, HD\,37594, HD\,49933
\end{keywords}
\section{Introduction}\label{introduction}
The advent of space missions aiming to obtain very accurate photometry for an
increasing number of stars (e.g. CoRoT and Kepler) led to the necessity of 
a large scale work to obtain high precision spectroscopic fundamental 
parameters, effective temperature in particular, to allow i.e. the modelling 
of the pulsation frequencies or the characterisation of transiting exoplanets. 
This large spectroscopic analysis campaign can be performed within reasonable 
time-scales only with automatic and semi-automatic procedures. It is 
therefore crucial to critically compare some of the results obtained in this 
way with other independent, more ``classical", methods and highlight the 
discrepancies to understand their origin and improve these procedures. 

\citet{bruntt-solar} derived fundamental parameters for a set of 23 
solar-type stars adopting various different techniques (e.g. asteroseismology, 
parallax, spectroscopy), concluding that purely spectroscopic methods lead to 
results comparable to the more robust ones, although small corrections might 
be necessary. It is important to carefully assess whether there are cases 
where the spectroscopy fails in recovering the correct set of parameters, and 
if such erroneous parameters are produced, we need to understand why and 
correct the methodology.

In this work we concentrate on a few discordant results obtained with the 
semi-automatic procedure developed by H. Bruntt \citep[see e.g.][]{bruntt02} 
and adopted by many authors to analyse several different types of stars.

For the solar-type pulsator HD\,49933, \citet{gillon} and \citet{bruntt} 
(hereafter B08), making use of two very similar semi-automatic 
procedures, derived an effective temperature (\Teff) of about 6750\,K. 
\citet[][hereafter R09]{ryabchikova2009} re-analysed this same star applying 
a ``classical" analysis (based on photometry, equivalent widths and hydrogen 
lines) to higher quality spectra, obtaining a \Teff\ of 6500\,K, 
demonstrating this is a better value for \Teff. Later \citet{bruntt09} 
reanalysed the same spectra of HD\,49933 used by R09 with the same method 
used by B08 obtaining a \Teff\ of 6570\,K, in agreement with R09, but in 
disagreement with B08. The later analysis did not explain the reason for such 
disparate results from the same semi-automatic procedure on the
same star, leaving open the possibility that different results might be 
obtained with data from different instruments. This possibility needs to 
be ruled out and it is important to understand the origin of these 
discrepancies, furthermore because the code employed by B08 and 
\citet{bruntt09} is adopted for most of the spectroscopic analysis of the 
CoRoT stars.

To understand the discrepancies, we must examine more than one star, therefore 
we decided to extend our critical analysis of B08's methodology 
for fundamental parameter determination to two other stars: HD\,32115 and 
HD\,37594, used as comparison stars by B08. \citet{bikmaev} showed the results 
of a fundamental parameter determination and abundance analysis of these two 
stars performed with a classical method, obtaining \Teff\ of 7250\,K and 
7170\,K, respectively. B08 re-analysed HD\,32115 and HD\,37594 employing a 
semi-automatic procedure on a different set of spectra, obtaining 7670\,K 
and 7380\,K, respectively. Here we use higher quality data than that 
previously used by \citet{bikmaev} and B08, focusing on the parameter 
determination and using only carefully selected spectral lines considering 
their \loggf\ values, blending, and known \nlte\ effects affecting those 
lines.

HD\,32115 is a single line spectroscopic binary for which \citet{fekel} 
determined an orbital period of about 8\,days, and concluded that the companion
is either a late K- or an early M-type star. This ensures that the spectral 
lines of the primary are not affected by the companion at any wavelength and 
that it is also safe to perform a spectrophotometric analysis to derive the 
stellar parameters of the primary star (see Sect.~\ref{sed}). 

For HD\,32115, \citet{fekel} derived an effective temperature of 7251\,K and 
a surface gravity of 4.26. From the HIPPARCOS parallax and from their orbit 
analysis, \citet{fekel} derived a stellar mass and radius of 1.5\,\M\ and 
1.5$\pm$0.1\,\R, respectively. By means of stellar structure and evolution 
calculations, \citet{allende} derived an effective temperature of 7413\,K and 
a surface gravity of 4.29$\pm$0.04. They derived also a stellar mass of 
1.54$\pm$0.06\,\M\ and a stellar radius of 1.48$\pm$0.03\,\R. We notice 
that \citet{bikmaev}'s \Teff\ for HD\,32115 coincides with the effective 
temperatures obtained by \citet{fekel}. On the other hand, 
B08's \Teff\ does not agree with any of the previous determinations.
\section{Observations}\label{obs}
Spectra of HD\,32115 and of HD\,37594 were obtained on 2010, November 30th
with the Robert G. Tull Coud\'e Spectrograph (TS) attached 
at the 2.7-m telescope of McDonald Observatory. This is a cross-dispersed 
\'echelle spectrograph yielding a resolving power ($R$) of 60\,000 
for the configuration used here. The signal-to-noise ratio (S/N) per pixel at 
$\lambda\sim$5000\,\AA\ is 490 and 535, respectively for HD\,32115 and 
HD\,37594.

Bias and flat-field frames were obtained at the beginning of the night, 
and a Th-Ar spectrum, for wavelength calibration, was obtained between 
the two science spectra, which were reduced using the Image Reduction 
and Analysis Facility 
\citep[IRAF\footnote{IRAF {\tt (http://iraf.noao.edu/)} is distributed by the National Optical Astronomy Observatory, which is operated by the Association of Universities for Research in Astronomy (AURA) under cooperative agreement with the National Science Foundation.},][]{tody}. 
Each spectrum, normalised by fitting a low order polynome to carefully 
selected continuum points, covers the wavelength range 3633--10849\,\AA, with 
several gaps between the orders at wavelengths greater than 5880\,\AA.
Making use of the Th-Ar spectrum we ensured the stability of the resolution.
The CCD count rates were well below the saturation level, ensuring linearity 
and therefore no systematic differences between strong and weak lines. This 
was also confirmed by the complete absence of any correlation between line 
abundance and wavelength (see Sect.\ref{metal lines}).

The normalisation of the H$\gamma$ line was of crucial importance since we
adopted the profile fitting of the H$\gamma$ line wings as one of the 
temperature indicators. We were unable to use either H$\alpha$ and H$\beta$
because the former was not covered by our spectra and the latter was affected 
by a spectrograph defect, preventing the normalisation. We were able to 
perform a reliable normalisation of the H$\gamma$ line using the artificial 
flat-fielding technique described by \citet{barklem02}. This normalisation
procedure was already successfully applied to data obtained with this
spectrograph by \citet{kolenberg}.
\subsection{Comparing spectrographs}\label{comparison}
As mentioned in Sect.~\ref{introduction}, \citet{bruntt09} left open the
possibility that the discrepancies obtained for HD\,49933 with their 
previous results (B08) were caused by systematic differences in the 
observed spectra \citep[][analysed HARPS spectra, while B08 analysed CORALIE spectra, where both instruments operate at LaSilla, but on different telescopes]{bruntt09}. 
We checked if this is the case. This control is important, to remove the 
observed spectra as a possible source of systematic uncertainties and also 
to check the quality of the normalisation, as independently performed on 
spectra obtained with different instruments.

Here we compare the spectra obtained with TS and CORALIE (used by B08), 
which is an \'echelle spectrograph, reaching a resolution of $R\sim$50\,000, 
mounted on the 1.2-m Euler telescope in La\,Silla, Chile. Details of the 
CORALIE spectrograph and on the data reduction can be found in \citet{decat}.

\begin{figure}
\begin{center}
\includegraphics[width=85mm,clip]{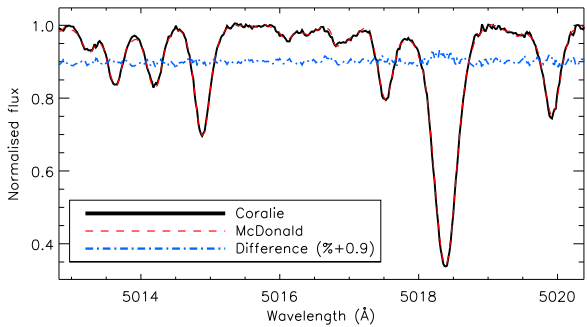}
\caption{Comparison between the spectra of HD\,32115 obtained with the 
CORALIE spectrograph (dashed red line) and the TS spectrograph (thick black
line) in the wavelength region around the $\lambda$5018\,\AA\ Fe\,{\sc ii} 
line. The difference (in \%) between the two spectra is shown by the dashed-
dotted blue line. The difference spectrum is shifted upwards by 0.9. No 
difference is visible between the two spectra, except for the core of the 
strong Fe\,{\sc ii} line at $\lambda$5018\,\AA\ which displays the difference 
in resolution between the two instruments.} 
\label{fig:comparison} 
\end{center} 
\end{figure}

Figure~\ref{fig:comparison} shows a comparison between the TS and the 
CORALIE spectrum of HD\,32115 in the wavelength range around the strong 
Fe\,{\sc ii} $\lambda$5018\,\AA\ line. In this plot we display also the 
difference spectrum (in \% shifted upwards of 0.9) between the two spectra.

The rms of the difference spectrum (a portion is shown in 
Fig.~\ref{fig:comparison}), calculated on several continuum regions is 
comparable to the rms of the CORALIE spectrum. We obtained the same 
conclusion comparing the TS and CORALIE spectra with the ones used by 
\citet{bikmaev}. The spectra of HD\,37594 demonstrate an identical behavior.

These comparisons let us conclude that there are no significant differences
between the spectra used in the present work and those employed by 
\citet{bikmaev} and by B08, thus excluding the quality of the data as the 
origin of the discrepancies described in Sect.~\ref{introduction}. 
In addition, this comparison sets an upper limit of $\sim$1\% on the 
uncertainty due to the normalisation.
\section{The fundamental parameters}\label{abn analysis}
We adopted photometric indicators to set the starting point in the 
determination of the fundamental parameters, which we refined making use 
of hydrogen lines, metal lines, and, as a final check, synthetic colors and 
the spectral energy distribution. Spectroscopic tuning of the fundamental 
parameters is needed because different photometric systems and calibrations 
give different parameters and uncertainties. Spectroscopic analysis,
performed with in this way, will produce a parameter set which fits all 
the indicators consistently, the uncertainties are thus reduced for those
derived from photometric analysis alone. 

We computed model atmospheres of HD\,32115 and HD\,37594 using the 
\llm\ stellar model atmosphere code \citep{llm}. For all the calculations 
Local Thermodynamical Equilibrium (LTE) and plane-parallel geometry were 
assumed. We used the \vald\ database \citep{vald1,vald2,T83av} as a source 
of atomic line parameters for opacity calculations. Convection was 
implemented according to the \citet{cm1,cm2} model of convection 
\citep[see][for more details]{heiter}. 
\subsection{Photometric indicators}
Initial guesses for the effective temperature (\Teff) and surface gravity 
(\logg) were obtained from calibrations of different photometric indices for 
normal stars. The effective temperature and gravity were derived from 
Str\"omgren photometry \citep{hauck} with calibrations by \citet{moon1985}, 
\citet{napiwotzki1993}, \citet{balona}, \citet{ribas}, \citet{castelli}, 
and from Geneva photometry \citep{rufener} with calibrations by 
\citet{north1990}.

Table~\ref{tab:phot} summarises the set of \Teff\ and \logg\ obtained with 
each adopted calibration and photometry, reflecting the scattering 
due to the use of different calibrations and photometric systems.
\input{./tables/photometry.tex}

The interstellar reddening plays an important role when converting the 
observed photometry into fundamental parameters. We calculated the 
interstellar reddening adopting the galactic extinction maps provided 
by \citet{ebv}, obtaining E($B-V$)=0.00 for both stars, which we 
adopt for the determination of the synthetic colors
(see Sect.~\ref{colors}) and spectral energy distribution 
(see Sect.~\ref{sed}). This is also in agreement with other models of
interstellar extinction in the solar neighborhood, e.g. \citet{lallement}.

Excluding the results obtained with the calibration by \citet{castelli}, which
gives much lower \Teff\ and \logg\ compared to the others, we set the center 
for the calculation of our model grid to \Teff\ = 7250\,K and \logg\ = 4.2, for
HD\,32115 and to \Teff\ = 7100\,K and \logg\ = 4.2, for HD\,37594, adopting
steps of 50\,K in \Teff\ and 0.1 in \logg.
\subsection{Spectroscopic indicators}
\subsubsection{Hydrogen lines}\label{hlines}
For a fully consistent abundance analysis, the photometric parameters must 
be checked and eventually tuned according to spectroscopic indicators, such as 
hydrogen line profiles. In the temperature range where HD\,32115 and
HD\,37594 lie, the hydrogen line wings are sensitive almost exclusively to 
\Teff\ variations. To spectroscopically derive \Teff\ from hydrogen lines, 
we fitted synthetic line profiles, calculated with \synth\ \citep{synth3}, 
to the observed H$\gamma$ profiles. \synth\ incorporates the code by 
\citet{barklem2000}\footnote{\tt http://www.astro.uu.se/$\sim$barklem/hlinop.html} 
that takes into account not only self-broadening but also Stark 
broadening (see their Sect.~3). For the latter, the default mode of \synth,
adopted in this work, uses an improved and extended HLINOP routine 
\citep{kurucz93}.

\begin{figure}
\begin{center}
\includegraphics[width=85mm,clip]{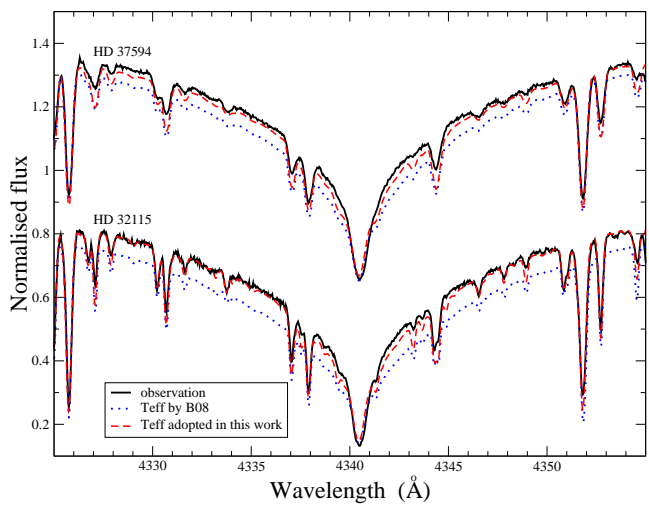}
\caption{Comparison between the observed H$\gamma$ line profile (black solid 
line) and synthetic profiles calculated with our final adopted parameters  
(red dashed line) and with the parameters adopted by B08 (blue dotted line).
The upper comparison is for HD\,37594, while the lower comparison is for
HD\,32115. Our final adopted set of parameters is \Teff\ = 7250$\pm$100\,K, 
\logg\ = 4.2$\pm$0.1 for HD\,32115 and \Teff\ = 7150$\pm$100\,K, 
\logg\ = 4.2$\pm$0.1 for HD\,37594. B08's set of parameters is 
\Teff\ = 7670$\pm$170\,K, \logg\ = 4.44$\pm$0.13 for HD\,32115 and 
\Teff\ = 7380$\pm$190\,K, \logg\ = 4.08$\pm$0.16 for HD\,37594. The profiles 
shown for HD\,37594 were rigidly shifted upwards by 0.5.} 
\label{hgamma} 
\end{center} 
\end{figure}

Figure~\ref{hgamma} shows a comparison of the observed H$\gamma$ 
line profiles of HD\,32115 and HD\,37594, with two synthetic profiles for each
star. One synthetic profile was calculated with our final set of parameters 
while the other one with the set of parameters by B08.

For both stars, the synthetic spectrum corresponding to our final model (see
Sect.~\ref{discussion}) does not fit the H$\gamma$ profile perfectly, this 
would require a lower temperature by $\sim$80--100\,K. We attribute 
most of the difference between our final synthetic and observed H$\gamma$ 
profiles to the normalisation, which is always challenging for hydrogen lines 
observed with \'echelle spectra. To quantify the uncertainty introduced by the 
normalisation, we compared the TS H$\gamma$ profile with the one we obtained 
from the CORALIE spectrum. The maximum difference between the two profiles 
(independently normalised) is $\sim$2\%, less than the difference introduced 
in the synthetic spectrum by changing \Teff\ by 100\,K.

Figure~\ref{hgamma} shows that the effective temperatures adopted by B08 for 
the two stars are too high to even remotely fit the hydrogen line profiles.
\subsubsection{Gravity from metallic lines with extended wings}
\label{magnisium}
The surface gravity was derived from two independent methods based on 
line profile fitting of Mg\,{\sc i} lines with developed wings (analysis 
described here) and ionisation balance for several elements (analysis 
performed in Sect.~\ref{metal lines}). The first method is described in 
\citet{fuhrmann} and is based on the fact that the wings of the Mg\,{\sc i} 
lines at $\lambda\lambda$\,5167, 5172 and 5183\,\AA\ are very sensitive to 
\logg\ variations. In practice, first the Mg abundance is determined from other 
Mg\,{\sc i} lines without developed wings, such as $\lambda$\,5528\,\AA, and 
then the wings of the three lines listed above are tuned to the \logg\ value.

It was not possible to use the Mg\,{\sc i} lines at $\lambda\lambda$ 5167 and 
5183\,\AA\ because the former is blended by a strong Fe\,{\sc i} line, while 
the latter falls at the border of two overlapping spectral orders and the 
normalisation was not good enough for the required precise fitting of the 
line wings. In the end, we based our fit for the \logg\ determination just on 
the Mg\,{\sc i} line at $\lambda$\,5172\,\AA.

For HD\,32115, we used the Mg\,{\sc i} line at $\lambda$\,5528\,\AA\ and the 
Mg\,{\sc ii} line at $\lambda$\,7877\,\AA\ to measure the Mg abundance, while 
for HD\,37594 we used only the former because the latter is blended with 
a telluric line, making the abundance determination unreliable. 

Given the presence of \nlte\ effects for Mg, in particular for the Mg\,{\sc ii} 
infrared line \citep{abia}, we calculated the \nlte\ corrections 
for the Mg lines used. As our synthetic line profiles are in LTE, \nlte\ 
calculations are necessary to make sure that the Mg abundance, we derive from 
the Mg\,{\sc i} $\lambda$\,5528\,\AA\ and Mg\,{\sc ii} $\lambda$\,7877\,\AA\ 
lines, is correct and therefore applicable for the line profile fitting of the 
Mg\,{\sc i} $\lambda$\,5172\,\AA\ line.

Non-LTE corrections for neutral and singly-ionised magnesium were carried out 
using the codes DETAIL and SURFACE, originally developed by \citet{giddings} 
and \citet{butler} along with the model atmosphere computed with \llm. Our 
calculations take into account the recent improvements in the atomic data 
for Mg, the extensive description of the model atom, and \nlte\ line 
formation presented by \citet{przybilla}. For both stars the \nlte\ 
abundance correction 
$\Delta(Mg/N_{tot}) = \log(Mg/N_{tot})^{LTE} - \log(Mg/N_{tot})^{NLTE}$ is 
$+0.02$\,dex for the Mg\,{\sc i} lines at $\lambda$\,5172 and 5528\,\AA, while 
for the Mg\,{\sc ii} infrared line at $\lambda$\,7877\,\AA\ we obtained a 
correction of $-0.07$\,dex, in agreement with the results by \citet{abia} 
and \citet{przybilla}.

In HD\,32115 for the Mg\,{\sc i} $\lambda$\,5528\,\AA\ line we obtained a LTE 
abundance of $\log(Mg/N_{tot})^{LTE}=-4.50$, therefore 
$\log(Mg/N_{tot})^{NLTE}=-4.48$. Similarly, for the Mg\,{\sc ii} line at 
$\lambda$\,7877\,\AA\ we obtained a LTE abundance of 
$\log(Mg/N_{tot})^{LTE}=-4.39$, therefore $\log(Mg/N_{tot})^{NLTE}=-4.46$. 
Spectral synthesis in the region of the weaker Mg\,{\sc ii} lines at 
$\lambda\lambda$\,4390 and 4427\,\AA, are not sensitive to \nlte\ effects, 
requires an abundance in close agreement with that derived from the 
Mg\,{\sc i} lines. Therefore to perform the fit of the line wings of the 
Mg\,{\sc i}\,$\lambda$\,5172\,\AA\ line, we set the Mg abundance for 
HD\,32115 at -4.48, in \nlte, and -4.50, in LTE.

For HD\,37954, \nlte\ corrections are similar to those in HD\,32115, 
in particular they are identical for both Mg\,{\sc i} lines at 
$\lambda$\,5528\,\AA\ and $\lambda$\,5172\,\AA. Therefore we applied 
for the fitting of the Mg\,{\sc i} $\lambda$\,5172\,\AA\ line wings the LTE 
Mg abundance derived from the $\lambda$\,5528\,\AA\ line: 
$\log(Mg/N_{tot})^{LTE}=-4.77$.

To derive \logg\ from the fit of the Mg\,{\sc i} lines with extended wings, 
very accurate \loggf\ values and Van der Waals 
(log\,$\mathbf{\gamma_{\rm Waals}}$) damping constants are needed. Two sets of \loggf\
laboratory measurements for the Mg\,{\sc i} triplet are available. The first 
one based on the lifetime laboratory measurements \citep{AZ} came from the 
\vald\ database, while the second set, based on lifetimes and branching ratio 
measurements was recently published by \citet{Aldenius}. The accuracy of this 
set of transition probabilities is $\sigma$(\loggf)$=\pm$0.04\,dex.
Van der Waals damping constants in \vald\ are calculated by 
\citet{barklem}. Another estimate of (log\,$\mathbf{\gamma_{\rm Waals}}$) 
was given by \citet{fuhrmann}, who derived 
log\,$\mathbf{\gamma_{\rm Waals}}=-7.42$ from the fitting of the solar lines 
using the \citet{AZ} oscillator strengths. \citet{fuhrmann} noted that Stark 
broadening does not practically influence the Mg\,{\sc i} line profiles in the 
solar spectrum and in Procyon, while it might be more significant in hotter
stars. For our analysis we employed Stark damping constant 
log\,$\mathbf{\gamma_{\rm Stark}}=-5.44$ for the Mg\,{\sc i} triplet and 
log\,$\mathbf{\gamma_{\rm Stark}}=-4.63$ for the Mg\,{\sc i} 
$\lambda$\,5528\,\AA\ line calculated by \citet{DS-B}. Our calculations show 
that the line profile of the latter line in the spectra of both HD\,32115 and 
HD~37594 is not sensitive to Stark and Van der Waals broadening effects.

First, we checked the atomic parameters on the NSO solar flux atlas 
\citep{NSO}. LTE synthetic spectrum calculations in the region of the 
Mg\,{\sc i} triplet and of the Mg\,{\sc i} $\lambda$\,5528\,\AA\ line were 
performed for three different models of the solar atmosphere: MARCS 
\citep{marcs}, MAFAGS \citep{mafags}, and the one calculated with the \llm\ 
code. We fit the extended wings, not the cores of these lines which are 
subject to \nlte\ effects. Using log\,$\mathbf{\gamma_{\rm Waals}}$ from 
\citet{barklem} for all lines and transition probabilities from 
\citet{Aldenius} we derived the following LTE Mg abundance in the solar 
atmosphere: $\log (Mg/N_{\rm tot})=-4.54$ (MARCS), $-4.55$ (MAFAGS), and 
$-4.51$ (\llm). It corresponds to 7.50, 7.49 and 7.53 in logarithmic scale 
where $\log(H)$=12.00. Our estimates are closer to the Mg meteoritic 
abundance of 7.53$\pm$0.01 \citep{Lodders}, then to the most recent 
published value of the Mg abundance of 7.60$\pm$0.04 in the solar 
photosphere \citep{Asplund}.     

For HD\,32115 and HD\,37594, careful fit of the Mg\,{\sc i} 
$\lambda$\,5172\,\AA\ line profile, calculated with the transition 
probabilities by \citet{Aldenius} and damping constants from \citet{barklem} 
and \citet{DS-B}, results in the final value of \logg=4.2$\pm$0.1, in good 
agreement also with the gravity estimates from the photometric calibrations. 
Error estimates include the claimed $\pm$30\,\% error in the Stark damping 
constant calculations and also the possible errors in other spectral line 
parameters. The \logg\ value of 4.2$\pm$0.1 is also in very good agreement 
with the results obtained by both \citet{fekel} and \citet{allende} for 
HD\,32115.

\begin{figure}
\begin{center}
\includegraphics[width=85mm,clip]{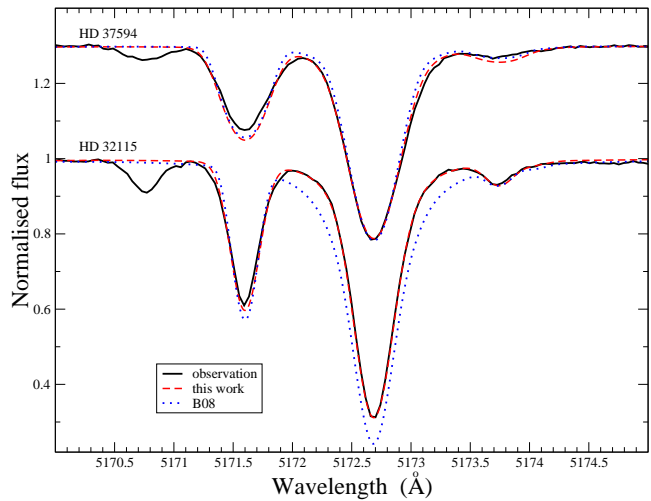}
\caption{Comparison between the observed Mg\,{\sc i} line profile of the line 
at $\lambda$\,5172\,\AA\ (black solid line) and synthetic profiles calculated 
with our final adopted parameters (red dashed line) and with the parameters 
adopted by B08 (blue dotted line). The upper comparison is for HD\,37594, 
while the lower comparison is for HD\,32115. The profiles shown for 
HD\,37594 were rigidly shifted upwards of 0.3. Our synthetic profiles are
calculated adopting the oscillator strength from \citet{Aldenius}, damping 
constant from \citet{barklem} and a LTE Mg abundance of -4.50\,dex for 
HD\,32115 and -4.77\,dex for HD\,37594. B08 does not provide a Mg abundance for 
HD\,37594, therefore we derived it from the Mg\,{\sc i} line at 
$\lambda$\,5528\,\AA, adopting the parameters given in their paper.} 
\label{mglines} 
\end{center} 
\end{figure}

Figure~\ref{mglines} shows comparisons for HD\,32115 and HD\,37594 between 
the observed and synthetic line profile of the 
$\lambda$\,5172\,\AA\ Mg\,{\sc i} line, calculated adopting the oscillator 
strength from \citet{Aldenius} and damping constant from \citet{barklem}. 
The synthetic profiles calculated with the parameters and abundances by 
B08 are also shown (B08 adopted oscillator strength and damping constant given
in \vald).

Figure~\ref{mglines} shows an excellent agreement between the synthetic 
profiles calculated with our final adopted parameters and the observed 
spectrum. On the other hand, the synthetic profiles calculated with B08's
parameters display a disagreement with the observations, in particular for
HD\,32115. For HD\,37594 the discrepancy is rather small: the effect of 
the higher \Teff\ is mostly compensated by the lower gravity.

The other two Mg\,{\sc i} lines of the triplet ($\lambda\lambda$\,5167 
and 5183\,\AA) will provide the same results we obtained with the
$\lambda$\,5172\,\AA\ Mg\,{\sc i} line, as tested by R09 with the Sun, Procyon
and HD\,49933.

For HD\,32115, we used the stellar mass and bolometric magnitude by 
\citet{fekel}, in addition to our \Teff, to derive \logg:
\begin{math}
\log(g/g_{\odot})=\log(\M)+4\cdot\log(\Teff/T_{eff,\odot})+0.4\cdot(M_{\mathrm{bol}}-4.75).
\end{math}
We obtained a \logg\ value of 4.24, in good agreement with our previous
estimation and with the results by \citet{fekel} and \citet{allende}.
\subsubsection{Metallic lines}\label{metal lines}
The metallic-line spectrum provides further constraints on the atmospheric 
parameters. If no deviation from LTE is expected, there should be no trend in
individual line abundances as a function of excitation potential. Examining 
this for any element/ion therefore provides a check on the determined value 
of \Teff. The balance between different ionisation stages of the same element 
similarly provides a check on \logg. The microturbulent velocity (\vmic) is 
subsequently determined by minimising any trend between individual abundances 
and equivalent widths for a certain ion. Determining the fundamental 
parameters in this way must be done iteratively since, for example, a 
variation in \Teff\ leads to a change in the best \logg\ and \vmic. This 
methodology of fundamental parameter determination from the metallic line 
spectrum is adopted in almost all semi-automatic abundance analysis 
procedures \citep[e.g.][]{bruntt02,santos04,gillon}.

The analysis of the metallic line spectrum requires the best possible knowledge
of atomic line parameters, \loggf\ values in particular. In this work we
only use lines of Ca, Ti, Cr, and Fe for which experimental atomic
parameters are available (except for Cr\,{\sc ii}, as clarified later in the 
text). Atomic parameters were extracted from the \vald\ database and from other 
recent publications, see references in Table~\ref{all-lines}. 

Data for neutral and ionised Ca lines were validated with \nlte\ calculations 
by \citet{Mashonkina2007}. For Ti\,{\sc i}, Ti\,{\sc ii}, Cr\,{\sc i} and 
Fe\,{\sc i} lines, accurate laboratory data (lifetimes and branching ratio) 
are available. For lines of ionised iron the oscillator strengths, selected 
from \vald, were produced from laboratory data, as explained in 
\citet{T83av}. All corresponding references are given in Table~\ref{all-lines}. 
Laboratory data for optical Cr\,{\sc ii} lines are 
scarce, therefore we took into account two different sets of calculated data, 
the one from semi-empirical orthogonal operator calculations by \citet{RU}, 
and the one from the latest calculations by R. Kurucz. The lifetimes 
calculated by the two groups agree very well and are very close to recent 
laboratory lifetime measurements by \citet{NLLN} and \citet{GNEL}. 
A small difference between theoretical and experimental lifetimes, converted 
to a difference in oscillator strength, corresponds to a \loggf-value 
uncertainty no larger than $\pm$0.03\,dex. However, the two sets of calculated 
oscillator strengths, for the lines used in our analysis, differ by 0.2\,dex, 
as a result of different branching factors, as the lifetimes are practically 
identical. From two theoretical sets of oscillator strengths, \citet{RU}'s 
data provide us with a smaller standard deviation from the mean Cr\,{\sc ii} 
abundance, while Kurucz's data give a mean abundance closer to that derived 
from the Cr\,{\sc i} lines. Clearly, an extensive laboratory analysis of 
Cr\,{\sc ii} lines in the optical region is needed for the 
interpretation of the Cr abundance in atmospheres of cool and hot stars. 

LTE line abundances were based on equivalent widths, analysed with a modified 
version \citep{vadim} of the WIDTH9 code \citep{kurucz1993a}. For 
blended lines and lines situated in the wings of the hydrogen lines we 
derived the line abundance performing synthetic spectrum calculations 
with the \synth\ code, tuning \vmac\ line by line (see 
Sect.~\ref{section:vmac}). A line-by-line abundance list with the equivalent 
width measurements, adopted oscillator strengths, and their sources is given 
in Table~\ref{all-lines} (see the online material for the complete version 
of the table). Table~\ref{all-lines} also gives equivalent width 
measurements for the lines which we measured via spectral synthesis.
In these cases, equivalent widths were tuned to match the abundance obtained 
with spectral synthesis. Our analysis shows that it is practically impossible 
to get a unique value of the microturbulent velocity for all considered 
species, therefore we derive a value that satisfies all the data and still 
provides a small scatter.
\input{./tables/alllines-short.tex}

\begin{figure}
\begin{center}
\includegraphics[width=85mm,clip]{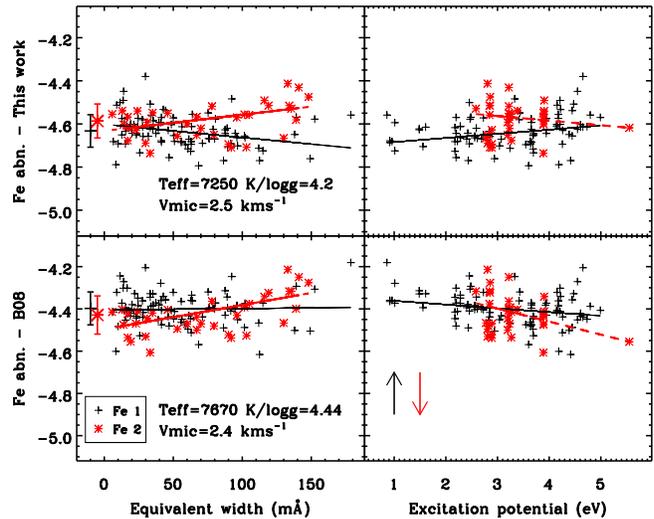}
\caption{Individual abundances for Fe\,{\sc i} (black crosses) and 
Fe\,{\sc ii} (red asterisks) lines versus the equivalent width (left panels) 
and the excitation energy of the lower level (right panels) for HD\,32115, 
with different model atmospheres. The top panels are for our adopted 
fundamental parameters, while the bottom panels are for B08's fundamental 
parameters. The linear fit to the Fe\,{\sc i} (black solid line) and 
Fe\,{\sc ii} (red dashed line) data are also shown. The black cross and red
asterisk at negative equivalent widths show respectively the mean Fe\,{\sc i} 
and Fe\,{\sc ii} abundance with their standard deviation. The black and red
arrows show the direction of the \nlte\ corrections for Fe\,{\sc i}
(upwards/higher abundance) and Fe\,{\sc ii} (downwards/lower abundance), 
respectively.} 
\label{fe_exc_hd32} 
\end{center} 
\end{figure}
\begin{figure}
\begin{center}
\includegraphics[width=85mm,clip]{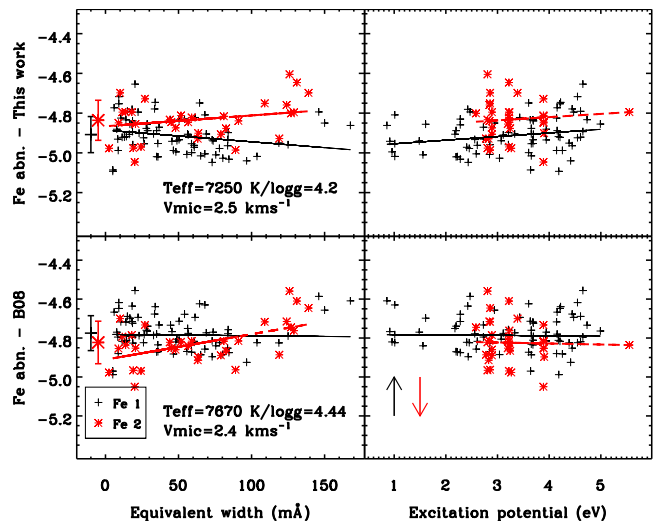}
\caption{Same as in Fig.~\ref{fe_exc_hd32}, but for HD\,37594.} 
\label{fe_exc_hd37} 
\end{center} 
\end{figure}

Figures~\ref{fe_exc_hd32} and \ref{fe_exc_hd37} show the correlations of 
Fe\,{\sc i} and Fe\,{\sc ii} abundance with equivalent width (left panels) and 
with excitation potential (right panels), respectively for HD\,32115 and 
HD\,37594. In each Figure, we used our final adopted fundamental parameters 
for the top panels and B08's fundamental parameters for the bottom panels. 
B08 derived the effective temperature by imposing the excitation
equilibrium for the Fe\,{\sc i} lines {\it only}, the surface gravity by 
imposing the ionisation equilibrium for Fe\,{\sc i} {\it only}, and the 
microturbulence velocity by removing the correlation between abundance and 
equivalent widths for Fe\,{\sc i} lines {\it only}.

With our model parameters for HD\,32115, we get a small positive 
correlation for Fe\,{\sc i} abundance, and a small anti-correlation for 
Fe\,{\sc ii} abundance, with the excitation potential. Our adopted \Teff\ 
thus accommodates both Fe\,{\sc i} and Fe\,{\sc ii}. With our model parameters 
for HD\,37594, we get a small positive correlation for both Fe\,{\sc i} and 
Fe\,{\sc ii} abundance with the excitation potential. These correlations 
would change slightly by adopting a different set of Fe lines, even by 
adding or removing a few lines, making the parameter determination based on 
these correlations rather sensitive to systematic effects introduced by the 
line selection. One example is the high excitation energy Fe\,{\sc ii} 
line, shown in Figs.~\ref{fe_exc_hd32} and \ref{fe_exc_hd37}, which has a 
much larger influence on the Fe\,{\sc ii} excitation equilibrium, compared 
to the other Fe\,{\sc ii} lines.

The model parameters obtained by B08 lead to almost perfect equilibria for 
Fe\,{\sc i}, as requested by B08 analysis method; in the case of HD\,32115,
B08's \Teff\ is anyway so high that even the excitation equilibrium for 
Fe\,{\sc i} is not reached, although it is imposed by their analysis method.

The equilibria for Fe\,{\sc i} are reached better with B08's parameters, 
compared to ours, as it is the {\it only} \Teff\ and \vmic\ indicator they 
used, but it leads to a set of parameters which does not fit all the other 
parameter indicators, first of all Fe\,{\sc ii}.

The average abundances, shown in Figs.~\ref{fe_exc_hd32} and \ref{fe_exc_hd37},
show that we obtain the ionisation equilibrium for Fe, within the error bars. 
We notice also that our average Fe\,{\sc ii} abundance is systematically higher
by a few 0.01\,dex, compared to Fe\,{\sc i}, in agreement with \nlte\ 
calculations by \citet{mashonkina} (the direction of the Fe \nlte\ corrections 
is shown in Figs.~\ref{fe_exc_hd32} and \ref{fe_exc_hd37}). Adopting 
B08's parameters, we obtain a systematically lower Fe\,{\sc ii} abundance, 
compared to Fe\,{\sc i}, and the \nlte\ corrections by \citet{mashonkina} 
would then worsen the ionisation equilibrium, indicating that B08's sets of 
parameters is inappropriate for these stars.

\begin{figure}
\begin{center}
\includegraphics[width=85mm,clip]{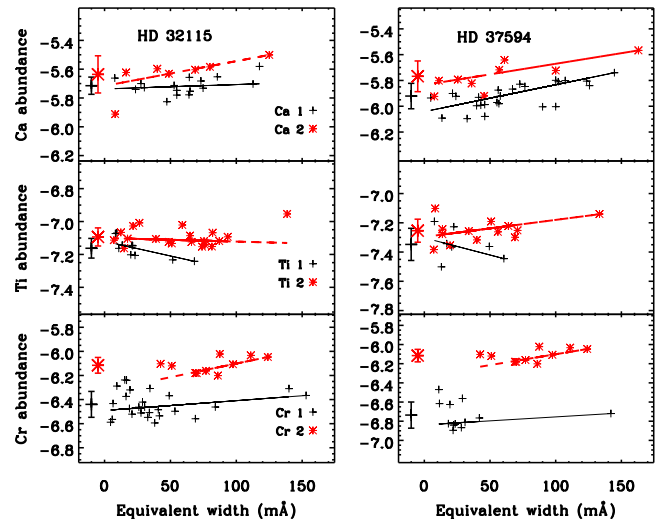}
\caption{Individual abundances for Ca (top panels), Ti (middle panels), and Cr
(bottom panels) lines versus the equivalent width measured for HD\,32115 
(left panels) and HD\,37594 (right panels). In each panel, the black crosses 
indicate the lines of neutral elements, while the red asterisks indicate the
lines of singly ionised elements. All abundances are derived assuming our 
adopted fundamental parameters. The linear fits to the data are also shown. 
The black cross and red asterisk at negative equivalent widths show the 
average abundance respectively for the neutral and singly ionised elements, 
with their standard deviation. The poor Cr ionisation equilibrium is most 
likely due to the poor quality line data.} 
\label{others_eqw_hds} 
\end{center} 
\end{figure}

Figure~\ref{others_eqw_hds} shows the line abundance versus the equivalent
widths for Ca, Ti, and Cr, in HD\,32115 and HD\,37594, adopting our final 
fundamental parameters. As for Fe, these elements display a variety 
of small correlations and on average our fundamental parameters are the ones 
which suit at best all considered ions. The ionisation equilibrium is reached
within the error bars for all elements considered here, except Cr, for which 
the equilibrium is obtained adopting Kurucz's oscillator strengths, as shown 
in Table~\ref{all-lines}. Also with laboratory data for Cr lines, we expect an
improvement in the Cr ionisation equilibrium.
%
\section{Discussion}\label{discussion}
For HD\,32115 our final adopted set of parameters is: 
\Teff\ = 7250$\pm$100\,K, \logg\ = 4.2$\pm$0.1 and \vmic\ = 2.5$\pm$0.2\,\kms. 
For HD\,37594 we derived \Teff\ = 7150$\pm$100\,K, \logg\ = 4.2$\pm$0.1 and 
\vmic\ = 2.6$\pm$0.2\,\kms. The measured projected rotational velocity (\vsini) 
is 8.3 and 17.0\,\kms, respectively for HD\,32115 and HD\,37594. The observed 
line profiles required also a substantial macroturbulence velocity (\vmac), 
generally between 8 and 10\,\kms\ for both stars 
(see Sect.~\ref{section:vmac} for more details). These high values of \vmac\ 
are in line with an extrapolation of the \vmac\-\Teff\ correlations given 
by \citet{valenti} and \citet{bruntt-solar}.

Our fundamental parameters are not perfect, by definition, but they provide, 
within the error bars, the best description of all available observables: 
photometric colors, hydrogen and metallic line profiles.
\subsection{Model fluxes and observed photometry}
Other methods which should always be used to check the fundamental parameters
obtained by spectroscopic means are: {\it i}) comparison of synthetic and 
observed optical colors; {\it ii}) comparison between photometry, calibrated 
in flux, and synthetic spectral energy distribution (SED). We performed
these comparisons for HD\,32115 and HD\,37594, aiming also to estimate the 
precision which can be reached with these two methods.
\subsubsection{Synthetic colors}\label{colors}
Table~\ref{tab:colors} summarises the comparison between observed and 
theoretical color-indexes for different photometric systems for the two stars, 
adopting the fundamental parameters obtained in this work and by B08.
We highlight the better agreeing theoretical index in each case. Overall out
values provide better matches. All colors were calculated using modified 
computer codes by \citet{kurucz1993a}, which take into account transmission 
curves of individual photometric filters, mirror reflectivity and a 
photomultiplier response function. In contrast to Kurucz's procedures 
\citep{relyea} which are based on the low resolution theoretical fluxes, 
our synthetic colors are computed from energy distributions sampled with 
a fine wavelength step, so integration errors are expected to be small.
\input{./tables/synt_phot.tex}

To fully understand Table~\ref{tab:colors}, it is first necessary to establish
more in general how well synthetic colors can reproduce the observations. 
This can be done by comparing synthetic and observed colors for 
well known stars. To check the quality of synthetic and observed colors over 
a large temperature range, we examine three ``standard" stars with 
different temperatures: Procyon \citep[\Teff=6530\,K;][]{fuhrmann}, 
Vega \citep[\Teff=9550\,K;][]{castellikurucz}, and 21\,Peg 
\citep[\Teff=10400\,K;][]{fossati2009}.

\begin{figure}
\begin{center}
\includegraphics[width=85mm,clip]{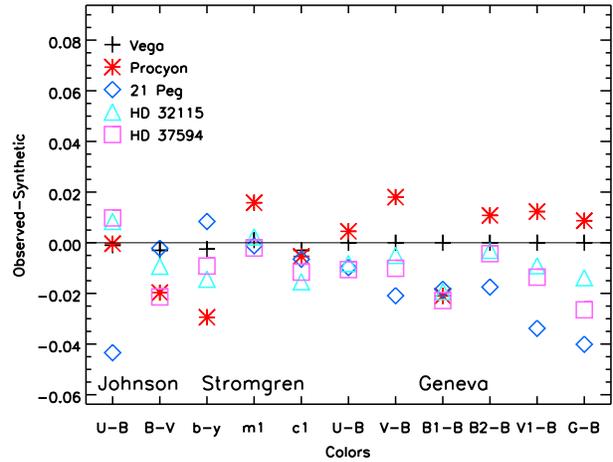}
\caption{Comparison between observed and synthetic colors for a set of 
reference stars, plus HD\,32115 (bright blue triangles) and HD\,37594 (violet
squares). The reference stars are: Vega (black crosses), Procyon (red
asterisks), and 21\,Peg (blue diamonds). Photometric colors are labelled on 
the X-axis. For all stars we assumed zero interstellar reddening.} 
\label{fig:colors} 
\end{center} 
\end{figure}

Figure~\ref{fig:colors} shows a comparison between synthetic and observed 
colors for the three reference stars, plus HD\,32115 and HD\,37594. With a few
exceptions, there is general good agreement between synthetic and observed 
colors throughout the temperature region explored here. For 21\,Peg, there 
is difference of 0.04\,mag between synthetic and observed $U-B$ Johnson color, 
$V_{\rm1}$-$B$ and $G$-$B$ Geneva colors. This difference is most likely due 
to incorrect photometry in one (or more) photometric band, as there is a 
perfect agreement between synthetic spectral energy distribution and 
spectrophotometry in the whole wavelength region between the near-UV and the
near infrared \citep[see Fig.~4 by][]{fossati2009}. On the other hand we do 
not have an explanation for the difference of 0.03\,mag obtained between 
synthetic and observed $b-y$ Str\"omgren color for Procyon, although an 
error in the photometry is always possible, even for such a bright star. 
Besides these two exceptions, there is general agreement between synthetic 
and observed colors at a 0.01--0.02\,mag level, which represents then the 
typical precision that one can expect for the comparison between observed and 
synthetic colors.

\begin{figure}
\begin{center}
\includegraphics[width=85mm,clip]{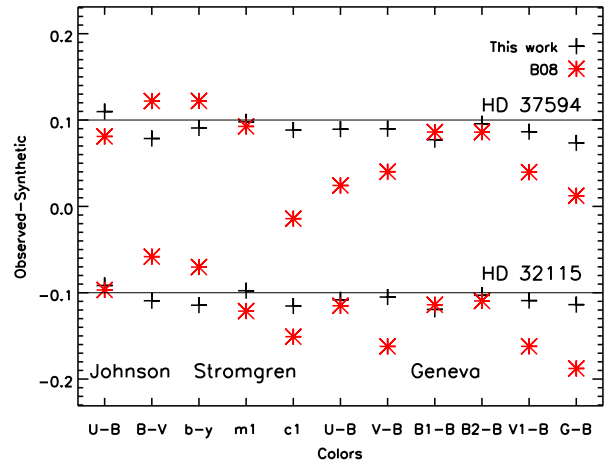}
\caption{Comparison between observed and synthetic colors for HD\,32115 
(shifted downwards of 0.1) and HD\,37594 (shifted upwards of 0.1). The colors 
calculated with our final set of parameters and B08's parameters are shown 
by black crosses and red asterisks, respectively. Photometric colors are 
labelled on the xaxis. For both stars we assumed zero interstellar 
reddening.} 
\label{fig:colors.comp} 
\end{center} 
\end{figure}

Figure~\ref{fig:colors} shows that for HD\,32115 and HD\,37594 the difference
between synthetic and observed colors is within the typical precision obtained
for the ``reference" stars, confirming the quality of our fundamental 
parameters. Figure~\ref{fig:colors.comp} shows that a comparison 
between observed and synthetic colors calculated with B08's stellar parameters 
clearly demonstrate the inadequacy of the B08 fundamental parameters.
%
\subsubsection{Spectral energy distribution}\label{sed}
For a complete self-consistent analysis of any star, one should reproduce 
the observed spectral energy distribution with the adopted parameters for a
model atmosphere. For HD\,32115 and HD\,37594 no spectrophotometry is 
available, therefore we converted the available Johnson \citep{johnson}, 
Geneva \citep{rufener} and 2MASS \citep{zacharias} photometry into physical 
units and compared them with the model fluxes calculated with the final 
set of parameters derived for the two stars. In the case of the Geneva colors, 
we assumed that Geneva $V_{\rm G}$ index is close to Johnson $V_{\rm J}$, 
making then possible to recover the Geneva $B$ index and thus $U$, $B_{\rm 1}$,
$B_{\rm 2}$, $V_{\rm 1}$, and $G$, and to transform them to absolute fluxes 
using the calibrations given by \citet{rufenernicolet}. For the 2MASS ($JHK$)
photometry we employed calibrations by \citet{vander}, while for
the Johnson photometry we employed calibrations by \citet{bessel}.

\begin{figure*}
\begin{center}
\includegraphics[width=170mm,clip]{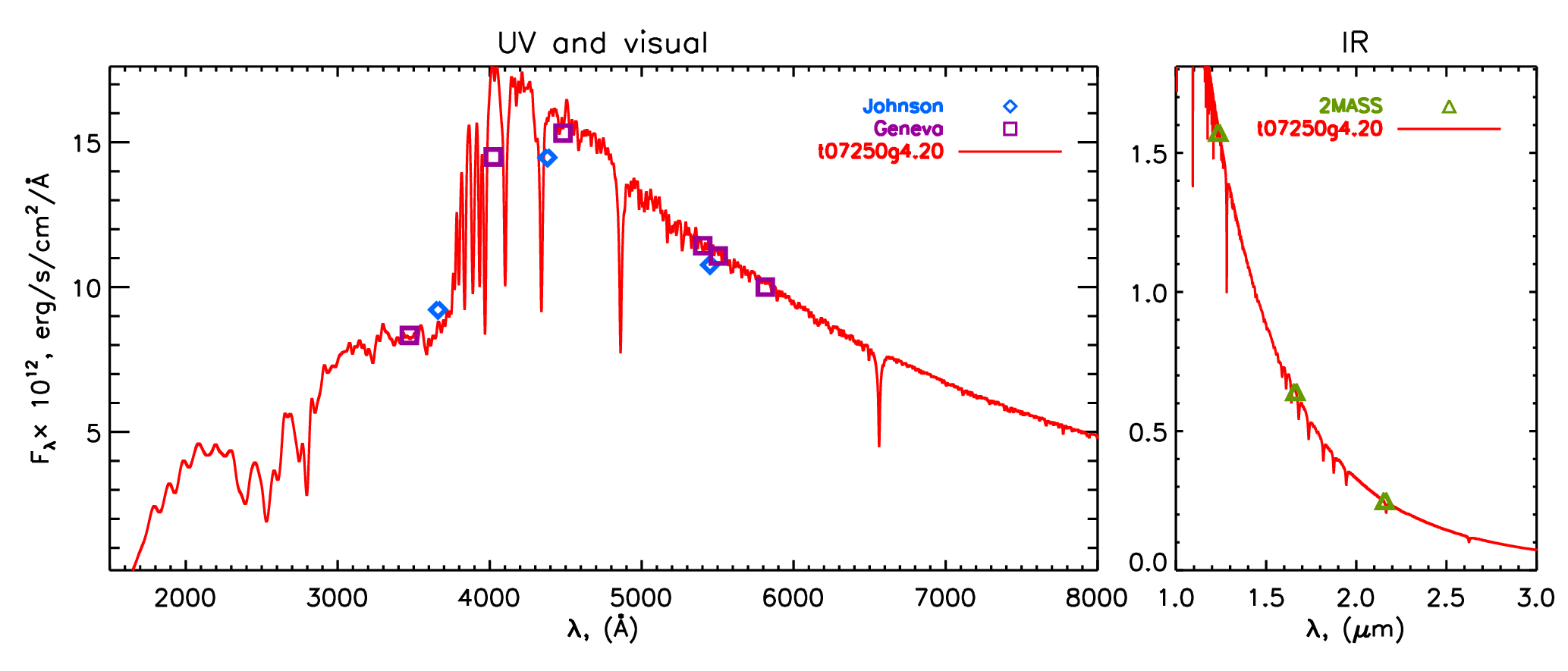}
\includegraphics[width=170mm,clip]{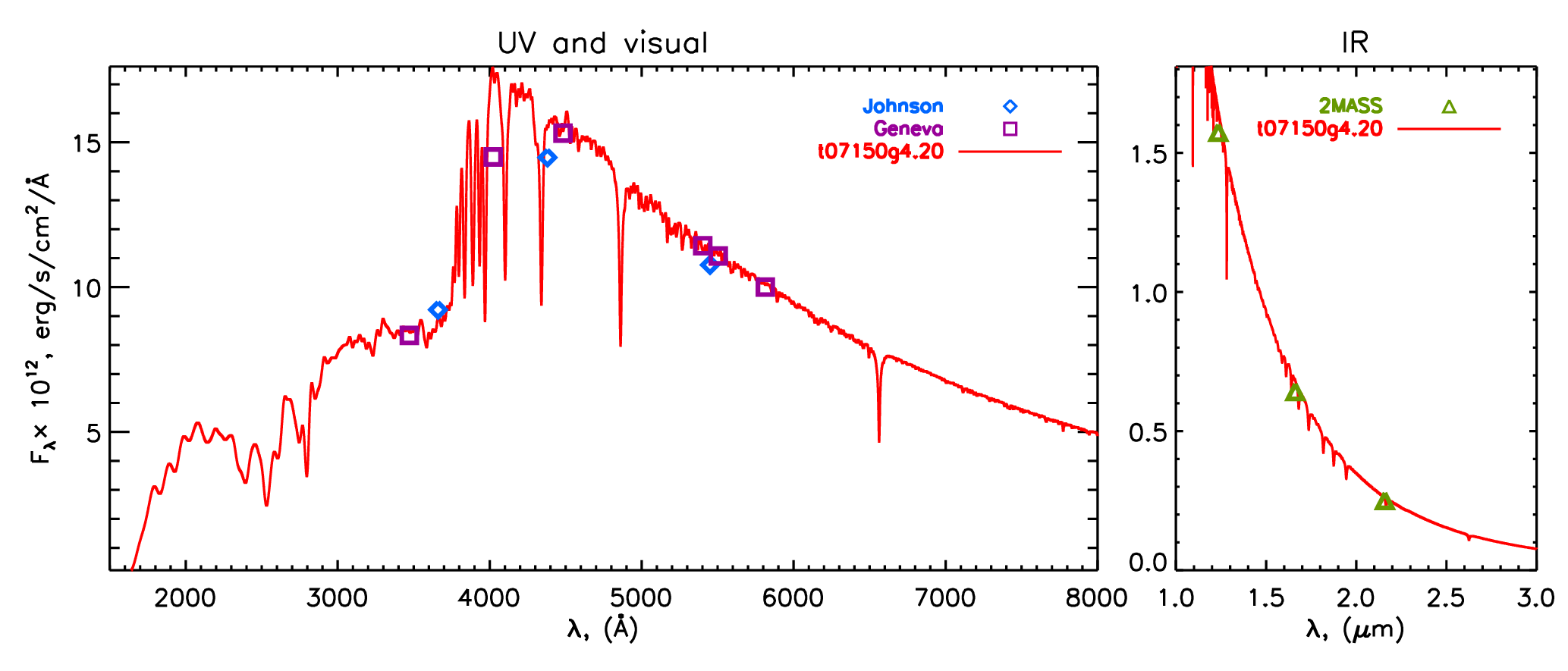}
\caption{Comparison between \llm\ theoretical fluxes (full red lines) 
calculated with the fundamental parameters derived for HD\,32115 (upper panel) 
and HD\,37594 (lower panel), with Johnson, Geneva (respectively open diamonds 
and squares in the left panels) and 2MASS (open triangles in the right panels) 
photometry converted to physical units.} 
\label{flux} 
\end{center} 
\end{figure*}

Figure~\ref{flux} shows the comparison between the observed photometry (in
physical units) and the model fluxes calculated with the adopted atmospheric 
parameters for HD\,32115 and HD\,37594. There is a good agreement between
observed and synthetic fluxes in the visible and infrared regions, 
providing a further confirmation of our adopted stellar parameters for 
the two stars.

Precise HIPPARCOS parallaxes are available for both stars \citep{leeuwen}.
This allowed us also to estimate their radii: 1.52$\pm$0.04\,$R_{\odot}$ 
for HD\,32115 and 1.36$\pm$0.02\,$R_{\odot}$ for HD\,37594. The radius we
derived for HD\,32115 is in very good agreement with what previously obtained 
by \citet{fekel} and \citet{allende}, who measured the stellar radius by 
means of stellar evolution calculations. This comparison increases the 
confidence on our results.
\subsection{Classic vs. semi-automatic procedures}\label{clavsauto}
The most evident difference between the analysis performed by B08 and our
analysis (including HD\,49933 by R09) is that B08 did not
add any \vmac\ to the line broadening. Since B08 used line profile fitting,
instead of equivalent widths, the lack of the additional broadening renders 
their whole analysis questionable. \citet{bikmaev} also did not use any 
\vmac\ in their analysis, but their spectra were of mid-low resolution and 
therefore \vsini\ was enough to fit the lines and their abundance analysis 
was based mainly on equivalent widths, independent of \vmac. At the 
resolution of CORALIE and TS ($R$=50\,000 and 60\,000, respectively), line 
profiles clearly require an additional \vmac\ broadening.
\begin{figure}
\begin{center}
\includegraphics[width=85mm,clip]{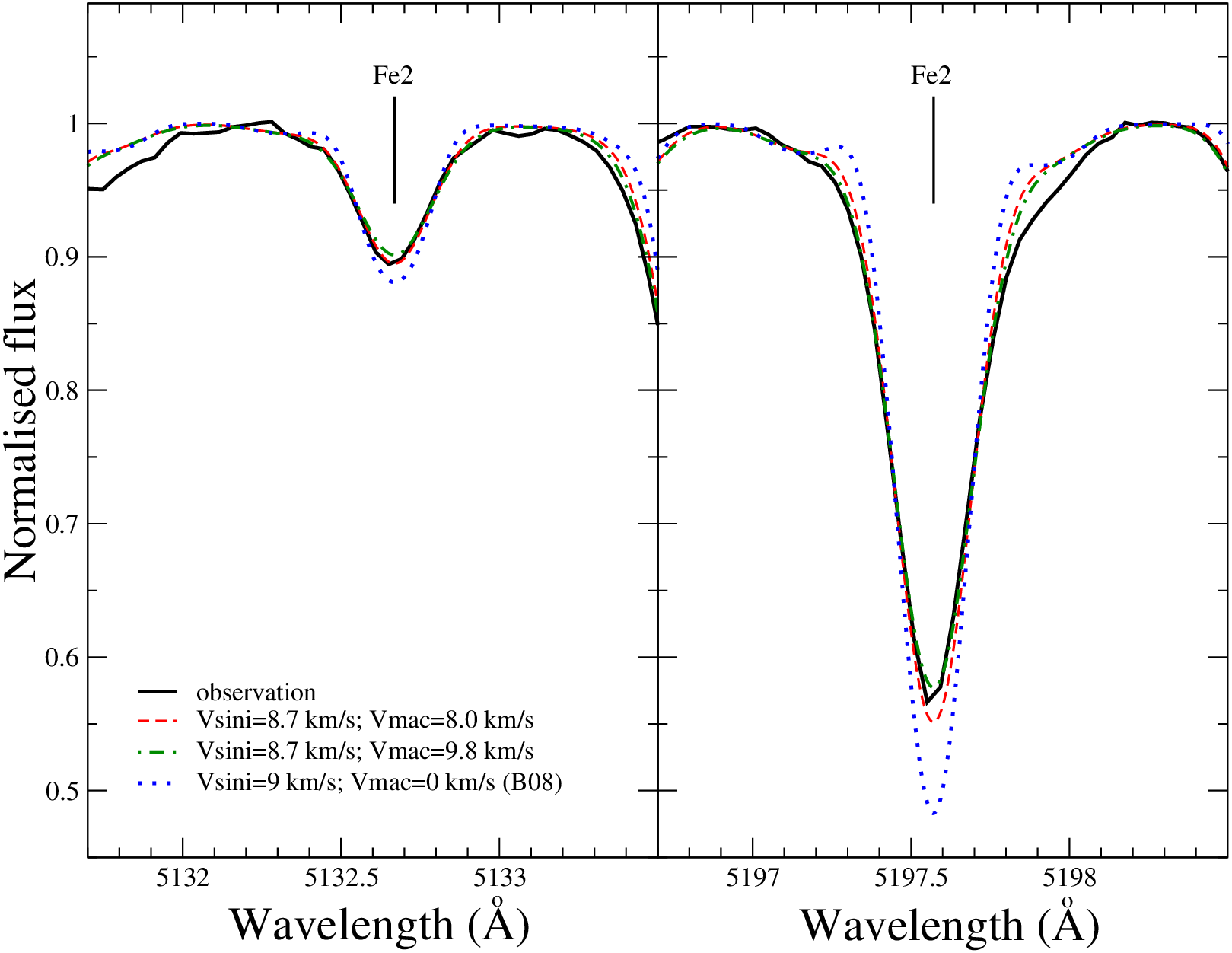}
\caption{Observed HD\,32115 line profiles (black solid line) of two 
Fe\,{\sc ii} lines at $\lambda$5132\,\AA\ (left panel) and 
$\lambda$5197\,\AA\ (right panel) in comparison with three synthetic spectra 
calculated with three different sets of broadening parameters: 
\vsini=8.7\,\kms\ and \vmac=8.0\,\kms\ (red dashed line); 
vsini=8.7\,\kms\ and \vmac=9.8\,\kms\ (green dotted-dashed line); 
\vsini=9.0\,\kms and \vmac=0\,\kms (blue dotted line). The first set of 
parameters fits the weak lines best, the second the strong lines, while the 
third is the one adopted by B08. For all three synthetic spectra we used our 
final set of fundamental parameters and abundances.} 
\label{fig:vmac} 
\end{center} 
\end{figure}

Figure~\ref{fig:vmac} compares the observed HD\,32115 line profiles of two 
Fe\,{\sc ii} lines, one weak and one strong, with three synthetic spectra 
calculated with our adopted parameters and three different sets of broadening 
parameters. The three broadening parameter sets are suited to weak lines, 
strong lines, and one as given by B08, respectively. 

The difference in \vmac\ between weak and strong lines is found systematically
for several lines of the same ion, and for different ions, excluding the
possibility that this is due to an error in \vmic, \vsini, or in the 
damping constants \citep[homogeneous calculations by][]{BPM,BA-J}. This
systematic difference in \vmac\ between weak and strong lines was found 
also by \citet{fuhrmann} in their analysis of Procyon.

The profiles calculated using the broadening by B08 show too deep 
cores and too narrow wings. This implies that B08 systematically obtained 
erroneously high \Teff\ for the three stars (HD\,49933, HD\,32115, and 
HD\,37594) to compensate for the deeper line cores, as a result of neglecting 
\vmac\ and fitting line cores to derive the abundances. This is confirmed by 
the fact that \citet{bruntt09} reanalysed HD\,49933, adopting a \vmac\ of 
2\,\kms, and obtained a much lower temperature than given in B08. It 
is still not clear to us where this particular value of \vmac\ came 
from, as R09 measured a \vmac\ of 5.2$\pm$0.5\,\kms, in 
agreement with the calibration by \citet{valenti}. Figure~\ref{fig:vmac} 
demonstrates that even adopting all the parameters and abundances given by 
B08, it is impossible to simultaneously fit the line profiles of weak and 
strong lines, as the wings will always be too narrow and the higher \Teff\ 
cannot compensate for it.

One of our main goals was to find and analyse possible sources of discrepancies 
between the results obtained with a classical method of analysis 
(e.g. this work and R09) and with a semi-automatic 
procedure \citep[e.g.][]{bruntt,bruntt09}. We have identified two possible
sources of discrepancy. The first arises from semi-automatic procedures 
often taking into account only Fe\,{\sc i} lines to derive \Teff\ and \vmic, 
while the second arises from the nature of the line profiles for these stars. 
\subsubsection{Is Fe\,{\sc i} alone good enough for a precise parameter
determination?}\label{fe1alone}
The upper-left panel of Fig.~\ref{fe_exc_hd32} shows the Fe\,{\sc i} and 
Fe\,{\sc ii} line abundance as a function of equivalent width, used for the 
determination of the microturbulence velocity. It is clear that a perfect 
equilibrium between the line abundance and equivalent width is not reached, 
and not simultaneously reachable, as the two ions show opposite correlations: 
$-5.524\pm2.407$ and $6.402\pm3.676$, in units of 10$^{-4}$, respectively 
for Fe\,{\sc i} and Fe\,{\sc ii}. Something similar is found for the ions of 
other species, as both Ca and Cr show a positive correlation, while Ti shows 
a negative correlation (see Fig.~\ref{others_eqw_hds}).

The \vmic\ value we adopted considers the effects on all the ions we took 
into account, whereas the \vmic\ value derived from the Fe\,{\sc i} lines 
alone, worsens the situation for Fe\,{\sc ii} and Ti, resulting in a 
systematic underestimation of \vmic. \citet{bruntt-solar} presented a 
calibration of \vmic\ as a function of \Teff\ for late-type stars, based on 
the analysis of Fe\,{\sc i} lines alone, therefore this calibration could 
also underestimate \vmic.

Using Fe\,{\sc i} lines alone for the determination of \vmic\ means 
that the average Fe\,{\sc ii} abundance depends strongly on which lines have 
been selected. If the selected Fe\,{\sc ii} lines are predominantly weak, the 
Fe\,{\sc ii} abundance will be artificially low. This introduces a bias in the 
determination of the surface gravity, as most semi-automatic procedures employ 
the Fe ionisation equilibrium alone to measure \logg. The use of predominant
medium-to-weak Fe\,{\sc ii} lines, with the adoption of Fe\,{\sc i} lines
alone to determine \vmic\ consequently leads to a sistematically erroneously 
low Fe\,{\sc ii} abundance, which then causes an erroneously high \logg\ to be
inferred. This of course, then affects the value of \Teff.

Another problem connected to the use of Fe\,{\sc i} lines as the sole 
temperature indicator is shown in the top-right panel of 
Fig.~\ref{fe_exc_hd32}. The effective temperature which best fits all 
available observables does not completely remove the correlation between 
abundance and excitation potential. For HD\,32115, by adopting only 
Fe\,{\sc i} as a \Teff\ and \vmic\ indicator, and only Fe for the ionisation 
equilibrium, we derived \Teff=7400\,K, \logg=4.2, and \vmic=2.4\,\kms. 
Similarly, for HD\,37594 we obtained \Teff=7300\,K, \logg=4.1, and 
\vmic=2.5\,\kms. These temperatures are too high and cannot fit the hydrogen 
lines or the photometry, with photometric colors revealing the
discrepancies.

The adoption of Fe\,{\sc i} alone as \Teff\ and \vmic\ indicator might also 
lead to the presence of systematic errors in the fundamental parameters arising
from \nlte\ effects. This is particularly true for cool stars, where \nlte\ 
effects are larger for Fe\,{\sc i} than Fe\,{\sc ii}, with obvious 
consequences for i.e. \logg. Non-LTE effects are generally stronger for 
stronger lines, implying that systematic errors might be introduced in the 
determination of \vmic, consequently affecting \Teff\ as well. 

All problems described here can be solved by including other ions in the
process of parameter determination. Although fewer lines of ions other than 
Fe\,{\sc i} are usually available, their inclusion in the parameter 
determination procedure would significantly alleviate the systematic errors 
introduced by the use of only Fe\,{\sc i} lines.

\citet{asplund05} shows differences between 3D and 1D LTE abundances for
several ions as a function of the excitation potential for the Sun. 
For Fe\,{\sc i}, the 1D models, compared to the 3D models, lead to a deviation 
as large as 0.2\,dex for the low excitation lines, which decreases with 
increasing excitation potential. Clearly, the use of 1D models introduces a 
strong bias towards high temperatures. This effect is present, with slightly 
different strengths, for many other ions. It would be extremely valuable to 
study this effect for stars hotter than the Sun, where the registered high 
\vmac\ values suggest that hydrodynamical effects might be even stronger 
than in the Sun.
\subsubsection{The effect of macroturbulence}\label{section:vmac}
\citet{valenti} and \citet{bruntt-solar} showed that for late-type stars the
macroturbulence velocity (\vmac) increases with increasing \Teff. For 
HD\,32115 and HD\,37594, \vmac\ reaches rather high values, close 
to 10\,\kms. During the analysis of the spectra of the two stars, we noticed 
a systematic difference between the line profiles of strong and weak lines, 
as strong lines require a systematically higher \vmac\ compared to weak lines, 
by $\sim$2\,\kms, as shown in Fig.~\ref{fig:vmac}. We believe that this 
difference is due to unmodelled depth-dependent velocities in the atmosphere 
which become evident for stars with relatively high \vmac\ and 
low \vsini.

This is not a concern if the abundances are derived from equivalent widths, as
they are independent of macroturbulence velocity. On the other hand, if the 
abundances are obtained from line profile fitting, where the abundance is 
the only free parameter, a systematic abundance difference is introduced 
between strong and weak lines, affecting the determination of the 
fundamental parameters. 

Semi-automatic abundance analysis procedures first set the broadening 
parameters to the entire available spectrum and then measure the line 
abundance from line profile fitting. If strong and weak lines require 
different broadening, opposing biases are introduced in the abundances 
derived from weak and strong lines. We estimated that the abundances 
obtained by fitting the profile of strong lines, adopting a \vmac\ typical 
for weak lines, are systematically lower by 0.10--0.15\,dex, compared to the 
abundances obtained with a \vmac\ appropriate to strong lines. Consequently 
the abundances obtained from weak and strong lines will be systematically 
over- and under-estimated, respectively. 

The systematic effect described above might be temperature and \vsini\ 
dependent as the difference in \vmac\ between weak and strong lines could
increase with increasing \Teff, and be more evident with decreasing \vsini.
In fact, the worst situation may be for slowly rotating ``cool"
early-type stars, such as HD\,32115.

The problem described here can be solved using equivalent widths, instead 
of line profile fitting, as the equivalent widths are broadening independent.
Alternatively \vmac\ could be treated as a further free parameter in the 
line profile fitting, although thorough tests on the uniqueness of the derived 
line abundance and \vmac\ should be performed. The adoption of 3D atmosphere
models would also partly eliminate this problem, but this is not 
a viable solution for an abundance analysis of a large sample of stars.
\section{Conclusion}\label{conclusion}
On the basis of high resolution, high signal-to-noise ratio spectra, taken 
at the McDonald observatory, we carried out a precise parameter 
determination of two late A-type stars, HD\,32115 and HD\,37594, which 
\citet{bruntt} adopted as reference stars for their abundance analysis 
of $\gamma$\,Dor stars. \citet{bruntt} analysed these stars with a 
semi-automatic procedure and their results strongly disagreed with those 
previously obtained by \citet{bikmaev}, by means of a ``classical" analysis. 
This discrepancy, together with that previously highlighted by 
\citet{ryabchikova2009} for HD\,49933, prompted us to reanalyse HD\,32115 
and HD\,37594 to look for the origin of the discrepancies.

We derived the fundamental parameters making use of all the available
observables: multicolor photometry, pressure-sensitive magnesium 
lines, metallic lines and profiles of hydrogen Balmer lines. For HD\,32115 
our final adopted set of parameters is: \Teff\ = 7250$\pm$100\,K, 
\logg\ = 4.2$\pm$0.1 and \vmic\ = 2.5$\pm$0.2\,\kms. For HD\,37594 we 
adopted: \Teff\ = 7150$\pm$100\,K, \logg\ = 4.2$\pm$0.1 and 
\vmic\ = 2.6$\pm$0.2\,\kms. We confirmed our final set of parameters by
comparing flux calibrated photometry with synthetic spectral energy
distributions, and by comparing observed and synthetic photometric colors.
Our prefered fundamental parameters fit, within the error bars, all available 
observables. They are also in agreement with the results by \citet{bikmaev}, 
but disagree with the results by \citet{bruntt}. For HD\,32115, our set 
of parameters agrees with that previously obtained by \citet{fekel} and 
\citet{allende}. In addition, the radius of HD\,32115, we derived from the 
analysis of the spectral energy distribution, is in perfect agreement with 
that derived by \citet{fekel} and \citet{allende} by means of stellar 
evolution calculations.

We compared our McDonald spectra with those used by \citet{bruntt} and 
\citet{bikmaev}, concluding that the differences between the three set of 
spectra is well within the given S/N. The quality of the data 
is not the origin of the discrepancies.  

To fit the line profiles of the two stars, we had to adopt rather large 
\vmac\ values, between 8 and 10\,\kms. Although the spectra analysed by 
\citet{bruntt} had enough resolution to require the need of \vmac\ to fit line
profiles, they adopted only rotational broadening (\vsini) in their spectral
synthesis. As a consequence, by deriving the line abundance by fitting 
synthetic line profiles to the observed ones, discarding any \vmac\ 
broadening, they introduced a bias in their analysis, which we believe 
led them to an erroneous set of fundamental parameters.

We have demonstrated that the determination of \Teff\ and \logg\ using only 
the Fe\,{\sc i} excitation equilibrium and the Fe ionisation equilibrium
leads to a systematic higher \Teff\ compared to that suggested by all other 
indicators. We also believe that this effect might be temperature 
dependent. 

Several automatic and semi-automatic procedures use the Fe\,{\sc i} 
excitation equilibrium and the Fe ionisation equilibrium as only/primary 
indicators for stellar parameter determination. In this work we show that 
these procedures do not always provide the correct set of fundamental 
parameters and their results need to be cautiously taken when performing 
further analysis, such as modelling of the asteroseismic frequencies 
or the characterisation of transiting exoplanets.
\section*{Acknowledgments}
Astronomy research at the Open University is supported by an STFC rolling 
grant (L.F., C.A.H.). D.S. is supported by Deutsche Forschungsgemeinschaft 
(DFG) Research Grant RE1664/7-1. KZ is recipient of an APART fellowship of 
the Austrian Academy of Sciences at the Institute of Astronomy of the 
University Vienna. We also acknowledge the use of cluster 
facilities at the Institute for Astronomy of the University of Vienna.
Data obtained with the 2.7-m telescope at McDonald Observatory, Texas, US. 
We thank Paul De Cat for providing us with the original CORALIE spectra.
%

\input{./tables/alllines.tex}


\bsp

\label{lastpage}

\end{document}

%% file: tables/photometry.tex
\begin{table*}
\caption[ ]{Set of \Teff\ and \logg\ obtained for HD\,32115 and HD\,37594 using 
different calibrations for Str\"omgren \citep{hauck} and Geneva \citep{rufener} photometry.  
The last two lines give the average values of \Teff\ and \logg, with their uncertainties, 
calculated without taking into account the results from the \citet{castelli} calibration.}
\label{tab:phot}
\begin{center}
\begin{tabular}{l|cccc|l}
\hline
\hline
 & \multicolumn{2}{c}{HD\,32115} & \multicolumn{2}{c}{HD\,37594} & \\ 
Photometry & \Teff~[K] & \logg & \Teff~[K] & \logg & Calibration \\
\hline       
Str\"omgren  & 7296 & 4.23 & 7171 & 4.22 & \citet{moon1985} \\
             & 7207 & 4.13 & 6899 & 3.86 & \citet{napiwotzki1993} \\
             & 7421 & 4.11 & 7276 & 4.05 & \citet{balona} \\
             & 7308 & 4.27 & 7189 & 4.32 & \citet{ribas} \\
             & 6998 & 3.84 & 6667 & 3.47 & \citet{castelli} \\
Geneva       & 7263 & 4.47 & 7157 & 4.47 & \citet{north1990} \\
\hline
 & 7300$\pm$80  & 4.24$\pm$0.14 & 7140$\pm$142 & 4.18$\pm$0.24 & \\
\hline							
\end{tabular}
\end{center}
\end{table*}

%% file: tables/alllines-short.tex
\begin{table*}
\caption{Lines used for the parameter determination. Wavelengths and 
excitation potentials are taken from the VALD database. The adopted \loggf\ values 
are taken from different sources which are listed in the last column. ``S" 
in the equivalent with column denotes line abundances determined by fitting the 
observed line profile, with the equivalent width determined from the line 
abundance. The \loggf\ values by \citet{BSScor} and 
\citet{BGHR} were corrected by +0.2 and +0.16, respectively. For Cr\,{\sc ii} 
the results obtained with two different sets of \loggf\ values are given 
(see the Sect.~\ref{metal lines}). For each ion, the last line gives the average
abundance and the standard deviation, with the number of lines in parenthesis.
The entire table can be viewed in the electronic version of the Journal.}
\label{all-lines}
\begin{center}
\begin{tabular}{ccl|cc|cc|r}
\hline
\hline
Element    &      &          & \multicolumn{2}{|c|}{HD\,37594} & \multicolumn{2}{c|}{HD\,32115} &            \\
Wavelength & \exc & \loggf~~ & EQW  & abundance                & EQW  & abundance	        & Ref \loggf \\
\AA        & eV   &          & m\AA & dex                      & m\AA & dex 		        &            \\
\hline   
Ca\,{\sc i} & & & & & & & \\
4425.4370 & 1.8790 & -0.358 & S 100.0  & -6.00 & 112.71 & -5.70 & SN \\
4435.6790 & 1.8860 & -0.517 & S  90.0  & -6.00 &        &       & SN \\      
4526.9280 & 2.7090 & -0.548 &          &       &  54.80 & -5.78 & SR \\      
4578.5510 & 2.5210 & -0.697 & 	 43.00 & -5.99 &  54.75 & -5.75 & SR+Sm	\\
4685.2680 & 2.9330 & -0.879 &  	 21.46 & -5.90 &  27.60 & -5.70 & S \\	
5261.7040 & 2.5210 & -0.579 & S  46.0  & -6.08 &        &       & SR+Sm \\   
5512.9800 & 2.9330 & -0.464 &	 39.77 & -6.00 &  47.26 & -5.83 & SR+Sm \\	
5581.9650 & 2.5230 & -0.555 &	 55.26 & -5.97 &  64.10 & -5.78 & SR+Sm \\	
5588.7490 & 2.5260 &  0.358 &   125.68 & -5.84 &	&	& SR+Sm \\   
...       & ...    & ...    &	 ...   & ...   &  ...   &  ...  & ... \\	
\hline
\end{tabular}
\end{center}
\smallskip

S  - \citet{S};\\
Sm  - \citet{Sm};\\
SN  - \citet{SN};\\
SR  - \citet{SR};\\
...
\end{table*}

%% file: tables/synt_phot.tex
\begin{table*}
\caption{Observed and calculated photometric parameters of HD\,32115 and HD\,37594. The values in brackets 
give the error bars of observations. The better agreeing theoretical index in each case is highlighted.}
\begin{center}
\begin{tabular}{lcccccc}
\hline\hline
 & \multicolumn{3}{c}{HD\,32115} & \multicolumn{3}{c}{HD\,37594} \\ 
Color           & Observed   & t7250g4.2 & t7670g4.44  & Observed   & t7150g4.20 & t7380g4.08 \\
index           & photometry & this work & B08         & photometry & this work  & B08        \\
\hline
\textbf{Johnson} &&&&\\
$U$-$B$         & 0.040  &     0.0316 &{\bf 0.0368}& 0.010  &{\bf 0.0002}& 0.0290	\\
$B$-$V$         & 0.285  &{\bf 0.2943}&     0.2432 & 0.275  &{\bf 0.2964}& 0.2530	\\ 
\\
\textbf{Str\"omgren} &&&&\\
$b$-$y$         & 0.176  &{\bf 0.1904}&     0.1463 & 0.189  &{\bf 0.1982}& 0.1668	\\ 
                & (0.003) &&& (0.001) &&\\
$m_{\rm 1}$     & 0.181  &{\bf 0.1787}&     0.2023 & 0.160  &{\bf 0.1621}& 0.1674	\\ 
                & (0.005) &&& (0.005) &&\\
$c_{\rm 1}$     & 0.689  &{\bf 0.7044}&     0.7399 & 0.669  &{\bf 0.6806}& 0.7832 	\\ 
                & (0.003) &&& (0.007) &&\\
$H\beta$        & 2.753  &{\bf 2.8077}&     2.8502 & 2.738  &{\bf 2.7924}& 2.8180	\\ 
                & (0.004) &&& (0.005) &&\\
\\
\textbf{Geneva} &&&&\\
$U$-$B$         & 1.389  &{\bf 1.3972}&     1.4041 & 1.354  &{\bf 1.3646}&     1.4298	\\ 
$V$-$B$         & 0.611  &{\bf 0.6160}&     0.6732 & 0.606  &{\bf 0.6161}&     0.6658	\\ 
$B_{\rm1}$-$B$  & 0.963  &     0.9823 &{\bf 0.9769}& 0.950  &     0.9728 &{\bf 0.9639}	\\ 
$B_{\rm2}$-$B$  & 1.416  &{\bf 1.4190}&     1.4255 & 1.423  &{\bf 1.4273}&     1.4371	\\ 
$V_{\rm1}$-$B$  & 1.328  &{\bf 1.3371}&     1.3901 & 1.324  &{\bf 1.3376}&     1.3844	\\ 
$G$-$B$         & 1.734  &{\bf 1.7478}&     1.8217 & 1.720  &{\bf 1.7465}&     1.8077	\\ 
\hline                                                                            
\end{tabular}                                                                     
\end{center}
\label{tab:colors}
\end{table*}

%% file: tables/alllines.tex
\onecolumn
\begin{longtable}{ccl|cc|cc|r}
\caption{Lines used for the parameter determination. Wavelengths and 
excitation potentials are taken from the VALD database. The adopted \loggf\ values 
are taken from different sources which are listed in the last column. ``S" 
in the equivalent with column denotes line abundances determined by fitting the 
observed line profile, with the equivalent width determined from the line 
abundance. The \loggf\ values by \citet{BSScor} and 
\citet{BGHR} were corrected by +0.2 and +0.16, respectively. For Cr\,{\sc ii} 
the results obtained with two different sets of \loggf\ values are given 
(see the Sect.~\ref{metal lines}). For each ion, the last line gives the average
abundance and the standard deviation, with the number of lines in parenthesis.}
\protect\label{all-lines}\\
\hline\hline \\
Element    &      &          & \multicolumn{2}{|c|}{HD\,37594} & \multicolumn{2}{c|}{HD\,32115} &            \\
Wavelength & \exc & \loggf~~ & EQW  & abundance                & EQW  & abundance	        & Ref \loggf \\
\AA        & eV   &          & m\AA & dex                      & m\AA & dex 		        &            \\
\hline   
\endfirsthead
\caption{continued.}\\
\hline\hline \\
Element    &      &          & \multicolumn{2}{|c|}{HD\,37594} & \multicolumn{2}{c|}{HD\,32115} &            \\
Wavelength & \exc & \loggf~~ & EQW  & abundance                & EQW  & abundance	        & Ref \loggf \\
\AA        & eV   &          & m\AA & dex                      & m\AA & dex 		        &            \\
\hline     
\endhead
\hline
\endfoot
\hline
\hline
\endlastfoot
Ca\,{\sc i} & & & & & & & \\
4425.4370 & 1.8790 & -0.358 & S 100.0  & -6.00 & 112.71 & -5.70 & SN \\
4435.6790 & 1.8860 & -0.517 & S  90.0  & -6.00 &        &       & SN \\      
4526.9280 & 2.7090 & -0.548 &          &       &  54.80 & -5.78 & SR \\      
4578.5510 & 2.5210 & -0.697 & 	 43.00 & -5.99 &  54.75 & -5.75 & SR+Sm	\\
4685.2680 & 2.9330 & -0.879 &  	 21.46 & -5.90 &  27.60 & -5.70 & S \\	
5261.7040 & 2.5210 & -0.579 & S  46.0  & -6.08 &        &       & SR+Sm \\   
5512.9800 & 2.9330 & -0.464 &	 39.77 & -6.00 &  47.26 & -5.83 & SR+Sm \\	
5581.9650 & 2.5230 & -0.555 &	 55.26 & -5.97 &  64.10 & -5.78 & SR+Sm \\	
5588.7490 & 2.5260 &  0.358 &   125.68 & -5.84 &	&	& SR+Sm \\   
5590.1140 & 2.5210 & -0.571 &	 57.83 & -5.92 &  64.85 & -5.75 & SR+Sm \\	
5601.2770 & 2.5260 & -0.523 &	 56.75 & -5.98 &  73.14 & -5.68 & SR+Sm \\	
5857.4510 & 2.9330 &  0.240 &   101.96 & -5.81 &	& 	& S+Sm \\   
5867.5620 & 2.9330 & -1.570 &	  4.87 & -5.94 &   7.73 & -5.66 & S \\	
6122.2170 & 1.8860 & -0.316 &   123.55 & -5.80 &	&	& SN \\      
6161.2970 & 2.5230 & -1.266 &	 13.42 & -6.09 &  23.49 & -5.74 & SR+Sm \\	
6162.1730 & 1.8990 & -0.090 &   144.67 & -5.74 &	&	& SN \\      
6166.4390 & 2.5210 & -1.142 &	 24.02 & -5.92 &  30.20 & -5.73 & SR+Sm \\	
6169.0420 & 2.5230 & -0.797 &	 32.57 & -6.10 &  52.80 & -5.71	& SR+Sm \\	
6169.5630 & 2.5260 & -0.478 &	 72.66 & -5.83 &  74.77 & -5.73	& SR+Sm \\
6471.6620 & 2.5260 & -0.686 &	 45.97 & -5.99 &  64.32 & -5.66	& SR+Sm	\\
6493.7810 & 2.5210 & -0.109 &	 99.99 & -5.80 &	&	& SR+Sm  \\  
6499.6500 & 2.5230 & -0.818 &	 41.29 & -5.93 &	&	& SR+Sm	\\   
6717.6810 & 2.7090 & -0.524 &	 56.24 & -5.87 &	&	& SR \\      
7148.1500 & 2.7090 &  0.137 &   107.26 & -5.80 & 117.58 & -5.58 & SR \\	
7202.2000 & 2.7090 & -0.262 &	 76.54 & -5.85 &	&	& SR \\      
7326.1450 & 2.9330 & -0.208 &	 67.13 & -5.87 &  85.61 & -5.65 & S \\	
\multicolumn{3}{c}{Average} & \multicolumn{2}{|c|}{-5.92$\pm$0.10 (25)} & \multicolumn{2}{|c|}{-5.71$\pm$0.06 (16)} & \\
\hline
Ca\,{\sc ii} & & & & & & & \\
5001.4790 & 7.5050 & -0.507 & S  36.00 & -5.82 &   48.88 & -5.63 & TB \\
5019.9710 & 7.5150 & -0.247 & S  45.50 & -5.92 &         & 	 & TB \\
5021.1380 & 7.5150 & -1.207 & S   7.60 & -5.92 &    8.00 & -5.91 & TB \\
5285.2660 & 7.5050 & -1.147 & S  11.10 & -5.80 & S 16.40 & -5.62 & TB \\
5339.1880 & 8.4380 & -0.079 & S  25.50 & -5.79 & S 40.00 & -5.60 & TB \\
6456.8750 & 8.4380 &  0.412 & S  61.00 & -5.64 &   80.00 & -5.58 & TB \\
8201.7220 & 7.5050 &  0.368 & S 100.00 & -5.72 &  125.03 & -5.50 & TB \\
8248.7960 & 7.5150 &  0.556 & S 163.00 & -5.57 &         & 	 & TB \\
8254.7210 & 7.5150 & -0.398 & S  57.00 & -5.72 &   68.79 & -5.60 & TB \\
\multicolumn{3}{c}{Average} & \multicolumn{2}{|c|}{-5.77$\pm$0.12 (9)} & \multicolumn{2}{|c|}{-5.64$\pm$0.13 (7)} & \\
\hline
Ti\,{\sc i} & & & & & & & \\
4453.6990 & 1.8730 & -0.010 &          &       &    9.30 & -7.07 & MFW \\
4548.7630 & 0.8260 & -0.354 &    17.07 & -7.34 &   20.80 & -7.14 & MFW \\
4617.2690 & 1.7490 &  0.389 &    13.00 & -7.50 &   19.90 & -7.20 & MFW \\
4758.1180 & 2.2490 &  0.425 &          &       &   10.80 & -7.16 & MFW \\
4759.2700 & 2.2560 &  0.514 &          &       &   13.38 & -7.14 & MFW \\
4981.7310 & 0.8480 &  0.504 &    60.55 & -7.44 &   68.20 & -7.24 & MFW \\
4999.5030 & 0.8260 &  0.250 &    49.47 & -7.36 &   51.70 & -7.23 & MFW \\
5192.9690 & 0.0210 & -1.006 &    22.40 & -7.23 &   21.30 & -7.15 & MFW \\
5210.3850 & 0.0480 & -0.884 &    20.84 & -7.36 &   23.10 & -7.21 & MFW \\
6261.0980 & 1.4300 & -0.479 &     7.67 & -7.19 &    8.20 & -7.07 & MFW \\
\multicolumn{3}{c}{Average} & \multicolumn{2}{|c|}{-7.35$\pm$0.11 (7)} & \multicolumn{2}{|c|}{-7.16$\pm$0.06 (10)} & \\
\hline
Ti\,{\sc ii} & & & & & & & \\
4394.0510 & 1.2210 & -1.770 &          &       &   93.50 & -7.09 & BHN \\
4395.8389 & 1.2430 & -1.970 &	       &       &   74.00 & -7.15 & BHN \\
4411.9250 & 1.2240 & -2.520 &	 33.86 & -7.26 &   39.00 & -7.11 & PTP \\
4417.7136 & 1.1650 & -1.190 &   133.34 & -7.14 &  138.58 & -6.95 & PTP \\
4418.3300 & 1.2370 & -1.990 &  	 70.91 & -7.25 &   75.50 & -7.12 & BHN \\
4421.9380 & 2.0610 & -1.660 &	 50.88 & -7.19 &   59.00 & -7.02 & PTP \\
4444.5545 & 1.1160 & -2.210 &	 63.98 & -7.22 &   66.00 & -7.12 & BHN \\
4470.8532 & 1.1650 & -2.020 &	       &       &   76.16 & -7.14 & PTP \\
4568.3140 & 1.2240 & -2.940 &	       &       &   21.63 & -7.03 & PTP \\
4609.2640 & 1.1800 & -3.430 &	  8.31 & -7.10 &    7.07 & -7.11 & BHN \\
4636.3200 & 1.1650 & -3.230 &	  7.18 & -7.38 &   12.35 & -7.07 & BHN \\
4657.2004 & 1.2430 & -2.320 &	       &       &   51.03 & -7.13 & BHN \\
4708.6621 & 1.2370 & -2.370 &	 40.00 & -7.32 &   48.70 & -7.12 & BHN \\
4779.9850 & 2.0480 & -1.260 &	       &       &   81.50 & -7.15 & RHL \\
5005.1570 & 1.5660 & -2.730 &	 13.78 & -7.28 &   17.44 & -7.10 & BHN \\
5185.9018 & 1.8930 & -1.490 &  S 69.0  & -7.30 &   82.00 & -7.07 & PTP \\
5252.0190 & 2.5900 & -1.960 &	 13.79 & -7.24 &   14.86 & -7.16 & PTP \\
5336.7710 & 1.5820 & -1.630 &	       &       &   87.09 & -7.12 & BHN \\
5381.0150 & 1.5660 & -1.970 &	 56.00 & -7.26 &   64.86 & -7.09 & BHN \\
7214.7160 & 2.5900 & -1.740 &	 19.60 & -7.35 &   26.44 & -7.01 & MFW \\
\multicolumn{3}{c}{Average} & \multicolumn{2}{|c|}{-7.25$\pm$0.08 (13)} & \multicolumn{2}{|c|}{-7.09$\pm$0.05 (20)} & \\
\hline
Cr\,{\sc i} & & & & & & & \\
4274.7970 & 0.0000 & -0.231 &  S 142.0 & -6.72 & S 153.0 & -6.37 & MFW \\
4545.9530 & 0.9410 & -1.370 &	 22.29 & -6.84 &  S 34.0 & -6.51 & SLS \\
4492.3050 & 3.3750 & -0.392 &	       &       &    9.32 & -6.29 & MFW \\
4616.1240 & 0.9830 & -1.190 &	 28.18 & -6.87 &   41.63 & -6.53 & SLS \\
4646.1620 & 1.0300 & -0.740 &	       &       &   68.94 & -6.56 & SLS \\
4651.2840 & 0.9830 & -1.460 &	 18.32 & -6.83 &   27.65 & -6.51 & SLS \\
4652.1570 & 1.0040 & -1.040 &	 41.90 & -6.76 &   53.42 & -6.50 & SLS \\
4689.3570 & 3.1250 & -0.400 &	 11.00 & -6.47 &   15.33 & -6.24 & SLS \\
4708.0130 & 3.1680 &  0.070 &	 19.50 & -6.62 &   20.79 & -6.52 & SLS \\
4718.4200 & 3.1950 &  0.240 &	 29.00 & -6.57 &   30.42 & -6.46 & SLS \\
4752.0870 & 4.1860 &  0.440 &	       &       &   16.68 & -6.24 & MFW \\
4756.1120 & 3.1040 &  0.090 &	       &       &   34.43 & -6.31 & MFW \\
4789.3350 & 2.5440 & -0.330 &	       &       &   28.78 & -6.42 & SLS \\
4922.2650 & 3.1040 &  0.380 &	       &       &   49.08 & -6.37 & SLS \\
4936.3360 & 3.1130 & -0.250 &	 11.50 & -6.61 &   16.30 & -6.37 & SLS \\
5206.0370 & 0.9410 &  0.020 &	       &       &  140.00 & -6.31 & SLS \\
5247.5650 & 0.9610 & -1.590 &	       &       &   26.00 & -6.46 & SLS \\
5296.6910 & 0.9830 & -1.360 &	 23.50 & -6.82 &   32.78 & -6.55 & SLS \\
5297.3770 & 2.9000 &  0.167 &	 21.91 & -6.89 &   40.75 & -6.48 & MFW \\
5348.3150 & 1.0040 & -1.210 &	 31.00 & -6.82 &   38.04 & -6.59 & SLS \\
5409.7840 & 1.0300 & -0.670 &	       &       &   84.09 & -6.46 & SLS \\
5783.0630 & 3.3230 & -0.500 &	       &       &    4.51 & -6.59 & MFW \\
5787.9180 & 3.3220 & -0.083 &	       &       &   19.30 & -6.32 & MFW \\
6925.2720 & 3.4490 & -0.330 &	       &       &    5.82 & -6.56 & MFW \\
6978.3970 & 3.4640 &  0.142 &	       &       &   18.87 & -6.47 & MFW \\
6979.7950 & 3.4640 & -0.410 &	       &       &    6.42 & -6.43 & MFW \\
\multicolumn{3}{c}{Average} & \multicolumn{2}{|c|}{-6.74$\pm$0.14 (12)} & \multicolumn{2}{|c|}{-6.44$\pm$0.11 (26)} & \\
\hline
Cr\,{\sc ii} & & & & & & & \\
4145.7810 & 5.3190 & -1.106 &          &       &  S 42.5 & -6.10 & RU \\
4252.6320 & 3.8580 & -2.054 &  S 33.0  & -6.44 &  S 51.0 & -6.12 & RU \\
4275.5670 & 3.8580 & -1.736 &  S 54.0  & -6.43 &  S 69.0 & -6.18 & RU \\
4554.9880 & 4.0710 & -1.491 &  S 56.0  & -6.50 &  S 77.0 & -6.16 & RU \\
4588.1990 & 4.0710 & -0.845 &  S 107.0 & -6.39 & S 124.0 & -6.05 & RU \\
4592.0490 & 4.0740 & -1.473 &  S 66.0  & -6.37 &  S 87.5 & -6.02 & RU \\
4616.6290 & 4.0720 & -1.576 &  S 56.0  & -6.45 &  S 70.0 & -6.18 & RU \\
4618.8030 & 4.0740 & -1.084 &  S 88.0  & -6.45 & S 111.0 & -6.03 & RU \\
4634.0700 & 4.0720 & -1.236 &  S 81.0  & -6.40 &  S 97.5 & -6.11 & RU \\
5237.3290 & 4.0730 & -1.350 &  S 71.0  & -6.45 &  S 86.0 & -6.20 & RU \\
\multicolumn{3}{c}{Average} & \multicolumn{2}{|c|}{-6.43$\pm$0.04 (9)} & \multicolumn{2}{|c|}{-6.12$\pm$0.07 (10)} & \\
4145.7810 & 5.3190 & -1.23  &	       &       &  S 42.5 & -5.98 & K10 \\
4252.6320 & 3.8580 & -1.99  &  S 33.0  & -6.50 &  S 51.0 & -6.18 & K10 \\
4275.5670 & 3.8580 & -1.67  &  S 54.0  & -6.50 &  S 69.0 & -6.25 & K10 \\
4554.9880 & 4.0710 & -1.28  &  S 56.0  & -6.71 &  S 77.0 & -6.37 & K10 \\
4588.1990 & 4.0710 & -0.63  &  S 107.0 & -6.61 & S 124.0 & -6.27 & K10 \\
4592.0490 & 4.0740 & -1.22  &  S 66.0  & -6.64 &  S 87.5 & -6.27 & K10 \\
4616.6290 & 4.0720 & -1.36  &  S 56.0  & -6.67 &  S 70.0 & -6.40 & K10 \\
4618.8030 & 4.0740 & -0.83  &  S 88.0  & -6.70 & S 111.0 & -6.28 & K10 \\
4634.0700 & 4.0720 & -1.02  &  S 81.0  & -6.62 &  S 97.5 & -6.33 & K10 \\
5237.3290 & 4.0730 & -1.14  &  S 71.0  & -6.66 &  S 86.0 & -6.41 & K10 \\
\multicolumn{3}{c}{Average} & \multicolumn{2}{|c|}{-6.62$\pm$0.08 (9)} & \multicolumn{2}{|c|}{-6.31$\pm$0.10 (10)} & \\
\hline
Fe\,{\sc i} & & & & & & & \\
4168.9416 & 3.4170 & -1.650 &	      &       &    27.60 & -4.58 & FMW \\
4189.5550 & 3.6940 & -1.330 &	      &       &    25.69 & -4.73 & FMW \\
4213.6474 & 2.8450 & -1.290 &	55.70 & -5.01 &    71.11 & -4.69 & FMW \\
4233.6020 & 2.4820 & -0.604 &	      &       &   139.12 & -4.57 & FMW \\
4248.2240 & 3.0710 & -1.286 &	48.90 & -4.94 &    61.29 & -4.67 & BWL \\
4250.1180 & 2.4690 & -0.405 &  146.00 & -4.79 &   156.31 & -4.58 & FMW \\
4266.9640 & 2.7270 & -1.812 &	36.67 & -4.87 &    49.96 & -4.57 & BWL \\
4267.8260 & 3.1110 & -1.174 &	56.61 & -4.91 &	         &	 & BWL \\
4433.2170 & 3.6540 & -0.700 &	58.90 & -4.95 &    80.53 & -4.58 & FMW \\
4484.2190 & 3.6020 & -0.864 &	64.40 & -4.75 &	         &	 & BWL \\
4485.6750 & 3.6860 & -1.020 &	35.83 & -4.95 &    50.21 & -4.65 & FMW \\
4602.0000 & 1.6080 & -3.154 &	16.70 & -4.86 &    21.98 & -4.61 & FMW \\
4602.9410 & 1.4850 & -2.209 &	73.70 & -4.94 &    87.52 & -4.63 & BWL \\
4630.1200 & 2.2790 & -2.587 &	18.00 & -4.86 &    24.27 & -4.61 & BWL \\
4635.8460 & 2.8450 & -2.358 &	      &       &    15.50 & -4.63 & BWL \\
4643.4630 & 3.6540 & -1.147 &	32.90 & -4.90 &    41.41 & -4.68 & BWL \\
4683.5597 & 2.8310 & -2.319 &	      &       &    15.29 & -4.69 & BWL \\
4690.1360 & 3.6860 & -1.645 &	14.00 & -4.83 &	         &	 & BWL \\
4733.5910 & 1.4850 & -2.988 &	      &       &    35.18 & -4.62 & BWL \\
4736.7720 & 3.2110 & -0.752 &	85.92 & -4.90 &    94.90 & -4.64 & BWL \\
4745.8000 & 3.6540 & -1.270 &	27.93 & -4.88 &    39.58 & -4.59 & BWL \\
4789.6508 & 3.5460 & -0.958 &	      &       &    58.77 & -4.70 & BWL \\
4962.5719 & 4.1780 & -1.182 &	17.00 & -4.90 &    24.49 & -4.57 & BWL \\
4966.0870 & 3.3320 & -0.871 &	70.10 & -4.90 &    92.31 & -4.48 & BWL \\
4994.1290 & 0.9150 & -3.080 &	48.70 & -4.94 &    55.75 & -4.69 & FMW \\
5014.9410 & 3.9430 & -0.303 &	76.40 & -4.93 &    87.60 & -4.64 & BWL \\
5044.2100 & 2.8510 & -2.038 &	18.70 & -4.96 &    28.09 & -4.66 & BWL,BK \\
5049.8190 & 2.2790 & -1.355 &	84.50 & -5.03 &   102.0  & -4.66 & BWL \\
5054.6420 & 3.6400 & -1.921 &	 5.00 & -5.09 &    10.0  & -4.69 & BWL \\
5083.3380 & 0.9580 & -2.958 &	48.20 & -4.99 &    62.86 & -4.68 & FMW \\
5090.7670 & 4.2560 & -0.400 &	46.49 & -4.99 &    63.00 & -4.69 & FMW \\
5127.3580 & 0.9150 & -3.307 &	      &       &    41.22 & -4.68 & FMW \\
5151.9100 & 1.0110 & -3.322 &	30.70 & -4.95 &    40.87 & -4.59 & FMW \\
5194.9410 & 1.5570 & -2.090 &	71.06 & -5.01 &    89.29 & -4.70 & FMW \\
5198.7110 & 2.2230 & -2.135 &	35.30 & -5.02 &    50.82 & -4.68 & FMW \\
5232.9390 & 2.9400 & -0.058 &  158.60 & -4.85 &   149.16 & -4.76 & BWL \\
5236.2020 & 4.1860 & -1.497 &	 5.00 & -5.09 &     8.38 & -4.79 & BWL \\
5242.4910 & 3.6340 & -0.967 &	50.40 & -4.83 &    55.67 & -4.68 & BWL \\
5269.5370 & 0.8590 & -1.321 &  167.67 & -4.86 &   178.64 & -4.56 & FMW \\
5281.7900 & 3.0380 & -0.834 &	80.31 & -5.03 &    96.56 & -4.70 & BWL \\
5283.6210 & 3.2410 & -0.432 &	      &       &   111.91 & -4.71 & BKK \\
5288.5247 & 3.6940 & -1.508 &	16.30 & -4.91 &    23.18 & -4.65 & BWL \\
5324.1780 & 3.2110 & -0.103 &  125.00 & -4.96 &	         &	 & BKK \\
5364.8580 & 4.4450 &  0.228 &	83.60 & -4.96 &   101.52 & -4.66 & BWL \\
5367.4660 & 4.4150 &  0.443 &	96.60 & -5.04 &   112.50 & -4.77 & BWL,BK \\
5373.6980 & 4.4730 & -0.860 &	22.30 & -4.79 &    33.26 & -4.51 & FMW \\
5379.5740 & 3.6940 & -1.514 &	18.51 & -4.84 &    23.33 & -4.64 & BWL \\
5386.3330 & 4.1540 & -1.770 &	 7.16 & -4.77 &     8.98 & -4.51 & FMW \\
5398.2770 & 4.4450 & -0.670 &	33.60 & -4.78 &    44.39 & -4.54 & FMW \\
5415.1920 & 4.3860 &  0.642 &  118.70 & -4.95 &   137.90 & -4.65 & BWL \\
5434.5230 & 1.0110 & -2.122 &  103.70 & -5.02 &   114.97 & -4.72 & FMW \\
5473.9000 & 4.1540 & -0.760 &	38.99 & -4.88 &    48.57 & -4.60 & FMW \\
5497.5160 & 1.0110 & -2.849 &	68.04 & -4.81 &	         &	 & FMW \\
5522.4460 & 4.2090 & -1.550 &	 9.10 & -4.75 &    13.71 & -4.49 & FMW \\
5576.0888 & 3.4300 & -1.000 &	61.30 & -4.82 &    75.26 & -4.55 & FMW \\
5618.6310 & 4.2090 & -1.276 &	11.40 & -4.92 &    19.03 & -4.60 & BWL \\
5633.9460 & 4.9910 & -0.270 &	28.50 & -4.85 &    41.42 & -4.57 & FMW \\
5638.2620 & 4.2200 & -0.870 &	34.02 & -4.75 &    42.79 & -4.53 & FMW \\
5662.5160 & 4.1780 & -0.573 &	      &       &    63.83 & -4.56 & BWL \\
5679.0229 & 4.6520 & -0.920 &	20.20 & -4.65 &    29.83 & -4.38 & FMW \\
5701.5440 & 2.5590 & -2.216 &	      &       &    31.85 & -4.66 & FMW \\
5753.1210 & 4.2600 & -0.688 &	      &       &    46.33 & -4.63 & BWL \\
5816.3730 & 4.5480 & -0.601 &	29.12 & -4.93 &    39.12 & -4.62 & FMW \\
5852.2170 & 4.5480 & -1.330 &	 9.10 & -4.72 &    14.00 & -4.45 & FMW \\
5905.6710 & 4.6520 & -0.730 &	19.42 & -4.90 &    27.80 & -4.61 & FMW \\
5916.2470 & 2.4530 & -2.994 &	      &       &    11.41 & -4.50 & FMW \\
5934.6530 & 3.9280 & -1.170 &	24.08 & -4.90 &    35.43 & -4.58 & FMW \\
6065.4820 & 2.6080 & -1.530 &	60.60 & -4.96 &    75.56 & -4.67 & FMW \\
6127.9060 & 4.1430 & -1.399 &	 8.70 & -4.99 &    16.49 & -4.61 & BWL \\
6136.6150 & 2.4530 & -1.400 &	86.11 & -4.98 &    91.65 & -4.69 & FMW \\
6136.9930 & 2.1980 & -2.950 &	      &       &    16.57 & -4.57 & FMW \\
6137.6910 & 2.5880 & -1.403 &	75.81 & -5.00 &    89.02 & -4.62 & FMW \\
6151.6170 & 2.1760 & -3.299 &	      &       &     6.51 & -4.68 & FMW \\
6165.3600 & 4.1430 & -1.474 &	10.50 & -4.82 &    14.46 & -4.60 & BWL \\
6180.2030 & 2.7270 & -2.586 &	 8.40 & -4.93 &    12.70 & -4.65 & BK \\
6187.9870 & 3.9430 & -1.720 &	 8.00 & -4.88 &    12.72 & -4.57 & FMW \\
6191.5580 & 2.4330 & -1.417 &	81.27 & -5.05 &    86.37 & -4.77 & BWL \\
6219.2790 & 2.1980 & -2.433 &	28.82 & -4.89 &    37.39 & -4.64 & FMW \\
6230.7220 & 2.5590 & -1.281 &	83.77 & -4.93 &    96.11 & -4.66 & FMW \\
6246.3170 & 3.6020 & -0.733 &	58.95 & -5.00 &    68.34 & -4.79 & BKK \\
6252.5540 & 2.4040 & -1.687 &	60.75 & -4.97 &    76.18 & -4.67 & FMW \\
6265.1310 & 2.1760 & -2.550 &	      &       &    33.28 & -4.62 & FMW \\
6335.3280 & 2.1980 & -2.177 &	35.89 & -5.02 &    46.94 & -4.76 & BWL \\
6336.8230 & 3.6860 & -0.856 &	50.37 & -4.93 &    60.57 & -4.72 & BK \\
6411.6470 & 3.6540 & -0.595 &	65.13 & -5.01 &    78.87 & -4.62 & BKK \\
6419.9420 & 4.7330 & -0.240 &	44.53 & -4.83 &    56.25 & -4.61 & FMW \\
6421.3490 & 2.2790 & -2.027 &	54.10 & -4.83 &    63.30 & -4.61 & FMW \\
6430.8440 & 2.1760 & -2.006 &	56.24 & -4.91 &    62.60 & -4.73 & FMW \\
6496.4660 & 4.7950 & -0.570 &	      &       &    33.86 & -4.56 & FMW \\
6677.9850 & 2.6920 & -1.418 &	      &       &    79.98 & -4.67 & BWL \\
6750.1500 & 2.4240 & -2.621 &	18.20 & -4.78 &    20.09 & -4.64 & FMW \\
6810.2570 & 4.6070 & -0.986 &	15.00 & -4.79 &    18.84 & -4.61 & BWL \\
6858.1450 & 4.6070 & -0.930 &	12.93 & -4.92 &    21.44 & -4.60 & BWL \\
\multicolumn{3}{c}{Average} & \multicolumn{2}{|c|}{-4.91$\pm$0.09 (75)} & \multicolumn{2}{|c|}{-4.63$\pm$0.07 (88)} & \\
\hline
Fe\,{\sc ii} & & & & & & & \\
4178.8620 & 2.5830 & -2.500 &S 138.9  & -4.80 & S 146.7  & -4.53 & T83av \\
4416.8300 & 2.7780 & -2.410 &  126.64 & -4.93 &   136.59 & -4.67 & BSScor \\
4491.4050 & 2.8560 & -2.700 &  116.13 & -4.75 &   127.01 & -4.51 & KK \\
4508.2880 & 2.8560 & -2.250 &  141.01 & -4.70 &   148.78 & -4.47 & T83av \\
4515.3390 & 2.8440 & -2.450 &  129.10 & -4.75 &   144.16 & -4.51 & T83av \\
4520.2240 & 2.8070 & -2.600 &  130.18 & -4.60 &   139.43 & -4.41 & T83av \\
4541.5240 & 2.8560 & -2.790 &   94.57 & -4.98 &   108.72 & -4.71 & BSScor \\
4576.3400 & 2.8440 & -2.920 &	92.81 & -4.85 &   106.02 & -4.56 & T83av \\
4620.5210 & 2.8280 & -3.240 &	63.16 & -4.93 &    79.48 & -4.65 & T83av \\
4731.4530 & 2.8910 & -3.000 & S 83.0  & -4.91 &    94.20 & -4.69 & BGHR \\
4993.3430 & 2.8070 & -3.640 & S 50.0  & -4.83 & S  52.0  & -4.66 & T83av \\
5132.6690 & 2.8070 & -3.980 &	20.63 & -4.98 &    29.97 & -4.69 & BSScor \\
5197.5770 & 3.2300 & -2.100 &  133.73 & -4.80 & S 141.0  & -4.58 & KK \\
5234.6250 & 3.2210 & -2.230 &  133.14 & -4.65 &   143.70 & -4.43 & HLGN  \\
5264.8120 & 3.2300 & -3.120 &	56.62 & -4.85 & S  74.9  & -4.62 & T83av \\
5284.1090 & 2.8910 & -2.990 &         &       &    94.64 & -4.71 & BSScor \\
5325.5530 & 3.2210 & -3.120 & S 66.6  & -4.82 & S  82.6  & -4.52 & BSScor \\
5414.0730 & 3.2210 & -3.540 & S 24.3  & -4.96 & S  48.6  & -4.55 & T83av \\
5425.2570 & 3.1990 & -3.160 &	      &       &    68.70 & -4.65 & BSScor \\
5525.1250 & 3.2670 & -3.950 &	      &       &  S 16.4  & -4.68 & HLGN \\
5534.8470 & 3.2450 & -2.730 &	82.53 & -4.88 &   101.58 & -4.56 & BSScor \\
5591.3680 & 3.2670 & -4.590 & S   2.8 & -4.98 &	    5.71 & -4.55 & RU \\
5627.4970 & 3.3870 & -4.130 & S 13.0  & -4.70 & S  15.0  & -4.54 & T83av \\
6084.1110 & 3.1990 & -3.780 &	21.80 & -4.85 &    31.07 & -4.59 & BSScor \\
6113.3220 & 3.2210 & -4.110 & S 13.7  & -4.79 & S  19.0  & -4.57 & BSScor \\
6149.2580 & 3.8890 & -2.720 &	51.68 & -4.82 & S  67.0  & -4.57 & BSScor \\
6247.5570 & 3.8920 & -2.310 &	89.16 & -4.82 &   106.52 & -4.57 & BSScor \\
6369.4620 & 2.8910 & -4.160 &	20.58 & -4.80 &    25.30 & -4.63 & BSScor \\
6383.7220 & 5.5530 & -2.210 &	13.74 & -4.80 &    19.23 & -4.62 & M \\
6416.9190 & 3.8920 & -2.790 &	50.76 & -4.85 & S  63.6  & -4.60 & M \\
6432.6800 & 2.8910 & -3.520 & S 48.0  & -4.88 &	         &       & T83av \\
6456.3830 & 3.9030 & -2.100 &	      &       & S 120.0  & -4.49 & BSScor \\
7224.4870 & 3.8890 & -3.240 &	25.60 & -4.72 &    37.85 & -4.55 & T83av \\
7449.3350 & 3.8890 & -3.090 & S 21.6  & -5.05 &    33.22 & -4.74 & HLGN \\
7479.6930 & 3.8920 & -3.680 &	 8.92 & -4.85 &    14.04 & -4.60 & BSScor \\
7515.8310 & 3.9030 & -3.460 &	      &       &    26.44 & -4.54 & T83av \\
7711.7230 & 3.9030 & -2.500 &	62.34 & -4.90 &	         &       & T83av \\
\multicolumn{3}{c}{Average} & \multicolumn{2}{|c|}{-4.84$\pm$0.10 (32)} & \multicolumn{2}{|c|}{-4.59$\pm$0.08 (35)} & \\
\hline
\end{longtable}
\smallskip

\twocolumn \noindent
BGHR  - \citet{BGHR};\\
BHN  - \citet{BHN};\\
BK  - \citet{BK};\\  
BKK  - \citet{BKK};\\
BSScor  - \citet{BSScor};\\
BWL  - \citet{BWL};\\
FMW  - \citet{FMW};\\
HLGN  - \citet{HLGN};\\ 
K10  - {\tt http://kurucz.harvard.edu/atoms.html};\\
KK  - \citet{KK};\\ 
M  - \citet{M};\\
MFW  - \citet{MFW};\\
PTP  - \citet{PTP};\\
RHL  - \citet{RHL};\\
RU  - \citet{RU};\\
S  - \citet{S};\\
SLS  - \citet{SLS};\\
Sm  - \citet{Sm};\\
SN  - \citet{SN};\\
SR  - \citet{SR};\\
T83av  - \citet{T83av};\\ 
TB  - \citet{TB}